\documentclass[fleqn,11pt]{proc} 
\usepackage{setspace}
\usepackage{listings}
\usepackage{multirow}
\usepackage{multicol}
\usepackage{titlesec}
\usepackage[T1]{fontenc}
\usepackage{subfig}
\usepackage{graphicx}
\usepackage[titletoc,toc,title]{appendix}
\usepackage{hyperref} 
\hypersetup{hidelinks,colorlinks,breaklinks=true,urlcolor=blue,citecolor=black,linkcolor=blue,bookmarksopen=false,pdftitle={Title},pdfauthor={Author}}

\title{Graphical processing unit accelerated enumeration and exploration of model genotype-phenotype maps for protein folding.} 
\author{Stephanie J. Owen}

\begin{document}
\abstract{Evolution can be broadly described in terms of mutations of the genotype and the subsequent selection of the phenotype.  The full enumeration of a given genotype-phenotype (GP) map is therefore a powerful technique in examining evolutionary landscapes.  However,  because the number of genotypes typically grows exponentially with genome length, such calculations rapidly become intractable. Here I apply graphics processing unit (GPU) techniques to the hydrophobic-polar (HP) model for protein folding.  This GP map is a simple and well-studied model for the complex process of protein folding.  Prior studies on relatively small 2D and 3D lattices have been exclusively carried out using conventional central processing unit (CPU) approaches.  By using GPU techniques, I was able to reproduce the pioneering calculations of Li et al.\cite{li1996emergence} with a speed up of 580-700 fold over a CPU.  I was also able to perform the largest enumeration to date of the 6$\times$6 lattice.  These novel calculations provide evidence that a popular `plum-pudding' metaphor that suggests that phenotypes are disconnected in genotype space does not describe the data.  Instead a `spaghetti' metaphor of connected genotype networks may be more suitable. Furthermore, the data allows the relationships between designability and complexity within GP space to be explored. GPU approaches appear extremely well suited to GP mapping and the success of this work provides a promising introduction for its wider application in this field.}


\flushbottom 

\maketitle 


\thispagestyle{empty} 


%
%

\section{Introduction} 

In the 1890s, the Dutch botanist Hugo de Vries rediscovered and expanded upon Mendel's groundbreaking work on heredity\cite{de1900spaltungsgesetz,mendel1866versuche}. This work revolutionized the field of genetics. Within 20 years, Mendel's principles of inheritance had been applied convincingly to model organism \textit{Drosophilia melanogaster}\cite{morgan1919physical} and mathematician RA Fisher had begun to develop the mathematical framework of population genetics\cite{fisher1919xv} which incorporated Mendel's principles into evolutionary theory. Fisher's work was the first quantification of the distribution and propagation of heritable genetic information. Ever since these great shifts in approach and understanding, scientists across many disciplines have been fascinated by the relationship between the information that is inherited by, and the resulting properties of, an organism\cite{geer1962genotype,schwartz2001genotype}.  In 1911, Johannsen introduced the fundamental words \textbf{phenotype}, \textbf{genotype} and gene to biology\cite{johannsen2014genotype}, vital terminology for the discussion of the nature of evolution. Phenotype, a broad term, refers to a set of observable characteristics such as the functional shape of a protein fold or the function of a gene regulatory network. The corresponding genotypes are encoded in the DNA and can be coarse-grained to the amino acid composition of a protein or the set of interacting genes in a regulatory network\cite{ciliberti2007innovation}. Thus the genotype refers to the information that generates the phenotype. 

 In 1991, Alberch formalized the ideas of a mapping between genotype and phenotype\cite{alberch1991genes}, leading to the introduction of the term genotype-phenotype (GP) map. Since then this concept has proven to be very fruitful because a GP map details which phenotypes are accessible to which genotypes. The number of genotypes that map to a phenotype is called its designability\cite{li1996emergence}. The GP map also encodes the possible consequences of mutations to the genotype. For example, it is well known that many mutations are effectively neutral; they don't change the phenotype\cite{duret2008neutral}.  A quantitative measure of the frequency of these neutral mutations is the robustness; phenotype robustness is the average probability that a mutation to the phenotype is neutral i.e. maps back to the same phenotype\cite{masel2010gpmap}. 

Exploring these properties for GP maps requires the exhaustive enumeration of all genotypes and phenotypes in the model. This can prove difficult for two reasons. Firstly, for any biologically realistic system, the simulation of a GP map is computationally expensive.  Secondly, for most biological systems an accurate mathematical description of the GP map is not known.  It is because of this that despite their utility, studies have typically been limited to relatively simple model systems.  Examples of GP maps that have been explored include simplistic sequence-structure relationships of molecules such as RNA or proteins\cite{lipman1991modelling}. GP maps of higher level systems such as gene transcription networks\cite{izquierdo2008evolution} and gene-regulatory networks have also been explored\cite{ciliberti2007innovation}.  Even for these simple systems computational expense remains a critical bottleneck for progress.

It should be kept in mind that the environment also plays an important role in evolution.  The GP map is not a complete description of evolution but without starting somewhere, progress can't be made. In fact, a great deal of work has been done on the evolutionary implications of data from model GP maps\cite{wagner1996perspective,shackleton2000investigation,pigliucci2010genotype}. It is thought that the  interplay of properties such as designability and robustness are important for understanding broader principles such as evolvability; the ability of evolution to generate novel heritable phenotypes.  For example, higher designability means that a phenotype is more likely to arise by random selection of a genotype. Similarly, high robustness means that genotypes that are more likely to form a percolating neutral network (NN) of genotypes that all map onto the same phenotype.  A NN in turn allows a population to travel through evolutionary space  over time without losing functionality\cite{smith1970natural}, but at the same time increases the number of alternative phenotypes that may be mutationally accessible.  Thus NNs are of key importance in providing adaptability to a system\cite{van1999neutral,rendel2011adaptive}.  

\paragraph{Spaghetti vs plum pudding model}

One key question for GP maps is whether or not the neutral networks of two phenotypes can be connected by single point mutations.   One extreme is the `spaghetti bowl' or `spaghetti' model (Figure 1A)  where the NNs are widely distributed and so phenotypes can make contact with many others.  The other extreme is the `plum pudding' model (Figure 1B) where NNs are much more isolated from one another.  This distinction is important, because the `spaghetti' model is thought to make a system much more evolvable, since it can more easily access evolutionary novelty through neutral exploration of its NN.

\begin{figure}[h!]\centering 
\includegraphics[width=0.7\linewidth]{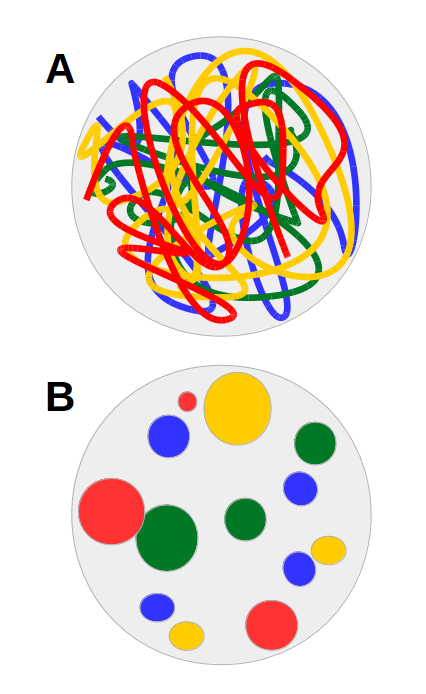}
\caption{Two hypothetical continuous genotype-phenotype spaces with different topologies; \textbf{A} depicts the `spaghetti bowl' model and \textbf{B} depicts the `plum pudding' model. Regions of color represent neutral networks, with the area representing the quantity of component genotypes.  Different colors represent different phenotypes. Gray represents deleterious phenotypes. For example in a GP map of protein folding, proteins that don't fold. In the `spaghetti' model, phenotypes percolate the space and make connections with many others. In the `plum pudding' model, phenotypes are isolated and make only limited contact with others.}
\label{fig:spag}
\end{figure}

\paragraph{Relationships between designability and complexity}

Biologically realistic GP maps typically have a small number of highly designable phenotypes that have most of the genotypes mapping to them\cite{schaper2014arrival,ferrada2012comparison}. Discovering what properties make some phenotypes exceptionally designable is an interesting problem in and of itself.  Such a discovery may also shed light on an explanation as to why only $10^3$-$10^4$ fundamentally different protein folds are observed in nature\cite{kolodny2013universe} whereas on theoretical grounds many orders of magnitude more are expected to be possible.
The properties of phenotypes that have so far been explored have been necessarily dependent upon the models used. Several authors have attempted to calculate what properties of phenotypes correlate with designability\cite{nochomovitz2006highly,wang2000symmetry}. An example of a property that has been investigated is the symmetry of protein folds in a 2D, lattice protein folding model. Li et al. noted when they looked at protein folding on a 6$\times$6 lattice that the highly designable phenotypes were highly symmetric\cite{li1996emergence}. Wang et al. took this further and investigated the links between the designability of protein folds on a $6\times6$ lattice and various symmetries of the folds\cite{wang2000symmetry} and found that for x-y symmetry and $180^{\circ}$ rotation symmetry, the designability of a fold increased with these symmetries on average.

Whilst attempting to find a model independent measure that correlates with designability, Louis et al. (unpublished) noted that highly symmetric structures may minimize a property called Kolgomorov complexity. Kolmogorov complexity is, put simply, the minimum length of a defining description of a object\cite{li2009introduction}. Highly symmetric structures include redundant information and as such may require a shorter description to specify. As such, this measurement of symmetry may be an indirect approximation of Komogorov complexity\cite{bonchev1976symmetry}. Further investigations into more complete measures of complexity may elucidate correlations between the designability of phenotypes and their complexity.

\paragraph{GP maps for protein folding models} There are 20 amino acids which make up the standard alphabet for protein sequences. Chains of amino acids spontaneously fold into complex, ordered structures called proteins. Modeling folding proteins has been of great theoretical interest since the classic experiment by Anfinsen in 1972\cite{anfinsen1973principles}. It was shown that for small globular proteins, all of the information a protein needs to fold is encoded in its amino acid sequence. 

\begin{figure}[h]\centering
\includegraphics[width=\linewidth]{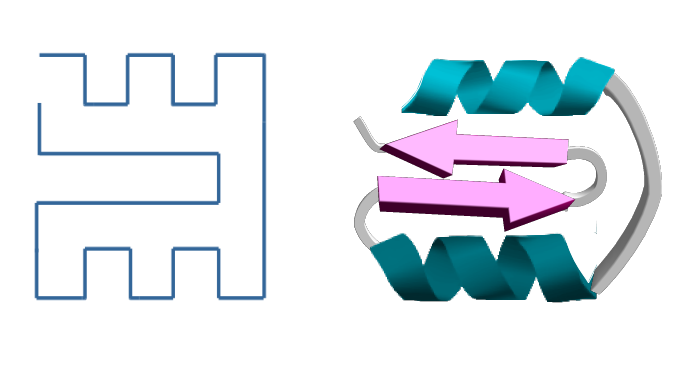}
\caption{An example of a path on a $6\times6$ lattice and a biologically similar structure with alpha helices (turquoise) and beta sheets (pink). This shows how 2D and 3D lattice structures can be translated into biological substructures.}
\label{fig:abfold}
\end{figure}

The protein folding problem is a grand challenge in science and many authors have approached the problem \cite{kallberg2012template,oldziej2005physics,wu2007ab}. Predicting the structure of any protein from its amino acid sequence is theoretically possible but practically impossible given current technology. The `Hydrophobic Polar' (HP) model is a much simplified protein folding model that attempts to tackle this complexity issue. The HP model was proposed by Dill in 1985\cite{dill1985theory} and exploits the fact that, to first order, all 20 amino acids can be separated into two categories: hydrophobic (H) and polar (P).   Secondly, the configuration space is greatly simplified to simple lattices.  The HP model therefore reduces a protein to a binary string of either polar or hydrophobic amino acids following a self-avoiding walk on a lattice.   It has been widely studied, and despite its simplicity, it is thought to reproduce some key aspects of the physics of protein folding.  A schematic showing how a 2D fold on a $6\times6$ lattice is similar to a biological fold comprised of alpha helices and beta sheets is shown in Figure \ref{fig:abfold}. 

It is more biologically realistic to calculate the solvation energy of protein folds with all 20 amino acids. This is typically done using the Miyazawa-Jernigan (MJ) energies, which are calculated statistically from known contacts within proteins\cite{miyazawa1985estimation}. The results of this model on the simple $3\times3\times3$ and $6\times6$ lattices have been compared to the simplified HP system; both models produce similar results, giving justification that the very simple and less computationally expensive HP model is biologically relevant\cite{miyazawa1985estimation}. 

The simplification provided by HP model is necessary in producing a GP map because calculating every genotype and phenotype for a model using all 20 amino acids is an intensive process; for a very modest protein of length 27 on a $3\times3\times3$ lattice, there are $20^{27} \approx 1 \times 10^{35}$ possible sequences and 103367 folds for which the energy of each sequence must be evaluated. Another example that illustrates the size of the genotype space for proteins is that all possible proteins of length 37 would weigh $1\times10^{25}kg$ which is approximately twice the mass of the Earth. When enumerating all sequences is impossible, the sampling of significant numbers of genotypes is often used to provide some insight into the structure of the GP space.

Although simplistic and therefore less computationally demanding than the MJ model, the HP folding problem is still NP complete\cite{berger1998protein}. This means that it is believed to be a computationally intractable problem. The solution space of this problem grows exponentially with its size\cite{mann2014exact} and there is no known shortcut to an exact solution\cite{berlekamp1978inherent}. In these models, even if we only look at a sample of all possible sequences, each sequence must still be exhaustively folded through all possible configurations in order to ensure the ground state is not missed. This is the primary reason why the acceleration of calculations is essential in being able to examine and create GP maps for systems in larger spaces than a $5\times5$ lattice.

The HP and MJ models therefore present a biologically relevant problem with a great need for acceleration and expansion. Several authors have derived properties of the GP map for these models on small lattices. In 2008, Goldstein investigated the HP model on the $5\times5$ lattice and suggested that the genotype-phenotype space for the HP model in general is `plum pudding' like\cite{Goldstein2008}. This would mean that the phenotypes are isolated, having limited connections with other `plums'. Ferrada and Wagner explored the same idea in 2012\cite{ferrada2012comparison}, suggesting as well that the phenotypes for proteins were less connected than the `highly interwoven' nature of the RNA GP map.

\paragraph{GPU computing applied to protein folding}
General purpose graphical processing unit computing is the use of graphical processing units (GPU) to accelerate programs that would be traditionally run on a central processing unit (CPU)\cite{luebke2004general}. This technique has seen a relatively recent uprising in usage for a variety of scientific problems\cite{owens2008gpu}. It achieved particular success in an Ising spin model (20-fold speed up)\cite{weigel2011simulating} and a non-hydrostatic weather model (80-fold speed up)\cite{shimokawabe201080}. The Folding@home project has been able to utilize consumer GPUs on laptops and home computers to achieve an average of 60-fold speed up on GPUs compared to CPUs of participants\cite{beberg2009folding}.

 This project brought together two exciting aspects of current research: 1) the analysis of biological genotype-phenotype spaces and 2) the use of GPUs to accelerate scientific simulations. 
 
 I wrote and verified a program that carried out the enumeration and sampling of HP and MJ models on given 2D and 3D lattices on the GPU. I was then able to investigate whether the HP GP space is more like `spaghetti' or `plum pudding' and whether the previous results from Ferrada and Goldstein were due to finite size effects. I then used HP and MJ models on the $6\times6$ lattice to investigate the link between various measurements of the information content or complexity of a phenotype and its designability. The minimum defining chain length and compression approximations of Kolmogorov complexity were evaluated for this lattice.

%
%

\section{Methods}
\subsection{HP and MJ lattice models}
The HP and MJ models use the H or P monomers or an alphabet of 20 amino acids respectively. These models are on lattice and require the full library of structures that a sequence can fold to. This library of possible folds is equivalent to all compact, self-avoiding random walks on the lattice. The use of only the compact structures is justifiable in that they quite accurately mimic globular proteins and that the most compact configuration of a protein is typically the most stable\cite{creighton1990protein}.
A stable protein fold is the lowest energy state for its sequence so finding the ground state is equivalent to minimizing the following Hamiltonian,

\begin{equation}
\label{eq:energy}
H =\displaystyle\sum\limits_{i<j} E_{\sigma_{i}\sigma_{j}}\Delta(r_i-r_j)
\end{equation}

\begin{figure}[h] \centering
\includegraphics[width=0.75\linewidth]{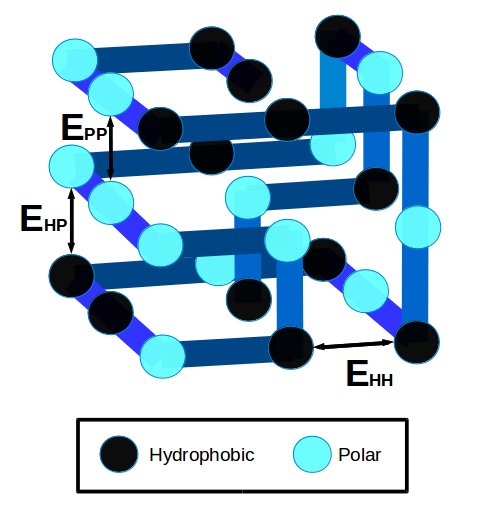}
\caption{The three different type of contact energies between neighboring H and P monomers on a $3\times3\times3$ lattice, $E_{\sigma_i\sigma_j}$ are defined in Equation 1 and Table 1.}
\label{fig:econtact}
\end{figure}

Where \textit{i} and \textit{j} label the position of the monomer in the sequence, $\sigma_i$ and $\sigma_j$ are the species of monomer (H or P in the HP model). $\Delta(r_i-r_j)=1$ for $i$ and $j$ which are neighbors in the fold but not neighbors in the sequence and $\Delta(r_i-r_j)=0$ for all other pairs. Examples of contacts that would contribute non-zero values to the Hamiltonian are shown in Figure \ref{fig:econtact} and the numerical values of $E_{\sigma_i\sigma_j}$ are given in Table \ref{tab:hp}. The values of $E_{\sigma_i\sigma_j}$ are from Li et al.\cite{li1996emergence} and the model using these energies will be consistently referred to as the HP model. The values of $E_{\sigma_i\sigma_j}$ for the MJ model are shown in Figure \ref{fig:MJmatrix} in Appendix A. 

\begin{table}[h]
\caption{HP model contact energies $E_{\sigma_{i}\sigma_{j}}$ in reduced units.}
\centering
\begin{tabular}{c c |c c}
\multicolumn{2}{}{} \\
& & $\sigma_i$  & \\
 & & H & P \\
\hline
$\sigma_j$ & H & $-2.3$ & $-1$ \\
& P & $-1$ & $0$ \\

\end{tabular}
\label{tab:hp}
\end{table}

The model takes the library of folds for a given lattice and then either samples or exhaustively enumerates sequences. For each sequence the energies for all configurations (folds) on the given lattice are found and the fold with the minimum energy is returned.
Each sequence in this model therefore has a lowest energy structure or it does not fold at all (i.e. has the deleterious phenotype). It is also possible to specify an energy threshold for further control over whether a sequence folds. The threshold is a parameter that sets the minimum distinguishable energy difference between the lowest and second lowest energy folds. If the two lowest energy configurations are within this threshold, there is no conclusive minimum and the sequence does not fold.

\subsection{Graphical Processing Units}
\label{sec:code}

\begin{figure}[h]\centering 

\includegraphics[width=\linewidth]{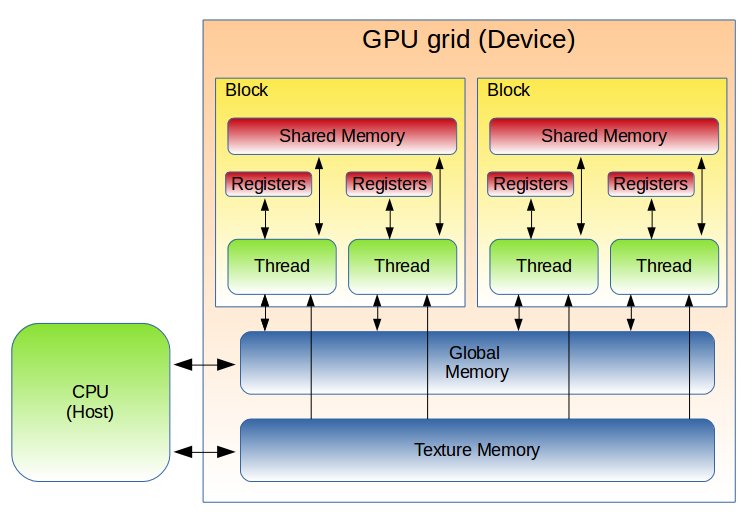}
\caption{Schematic of a GPU and its host CPU. The different components of the GPU are described in the main text.}
\label{fig:nvidia}
\end{figure}

GPUs are designed to carry out a large number of graphical calculations in parallel\cite{brodtkorb2013graphics} which is equivalent to carrying out many floating point matrix calculations. This infrastructure provides an excellent opportunity for the acceleration of scientific computing as it allows relatively simple operations to be carried out in parallel on cores (i.e. streaming multiprocessors) within the GPU using a system of blocks and threads (Figure \ref{fig:nvidia}). In 2007, Compute Unified Device Architecture (CUDA) was introduced by NVIDIA to simplify the process of using graphical processing units for general purpose computing\cite{10.1109/CGO.2007.13,nickolls2008scalable}. 

\begin{table}[h]
\caption{CUDA terminology}
\begin{tabular}{l   p{0.75\linewidth}} \hline \\
\textbf{Kernel}  & A function within the CUDA script, called by the host (CPU) and executed on the device (GPU) by an array of threads.\\
\textbf{Thread}   & A basic element of data to be processed, delivered to a single processing unit. \\
\textbf{Warp}   & A group of 32 threads. Threads are launched in warps.  One warp is scheduled on one multiprocessor.  \\
\textbf{Block}   & A group of threads that can cooperate via shared memory and synchronization. Should be a multiple of the warp size.\\
\textbf{Grid}   & A structure for launching multiple blocks.
\label{tab:cuda}
\end{tabular}
\end{table}

The memory within a computer with a GPU is organized into a hierarchy of 5 structures (in bold). Each multiprocessor includes thousands of \textbf{registers} which can be accessed locally and as such this memory access is very fast. Each multiprocessor has access to 48kB of \textbf{shared memory} which allows synchronization between threads in a block; this is the second fastest memory that can be accessed by a thread. There is approximately 4GB of \textbf{global memory} on separate DRAM chips that can be accessed by every thread but is the slowest memory access. \textbf{Texture memory} is a region of read-only memory that is specifically set aside for fast access and is available locally to each thread. The \textbf{host memory} is on the CPU and cannot be accessed by threads. Any memory transfers from the CPU to GPU must take place outside of running kernels. This memory hierarchy is illustrated in Figure \ref{fig:nvidia}.

Two different NVIDIA cards were used, the GeForce GTX 750Ti and Tesla C2070 cards which both use the Fermi architecture. The Intel® Core™ i5-480M CPU was used for all timed CPU calculations.

\paragraph{Compute Unified Device Architecture (CUDA) and the design of GPU code.}
CUDA was chosen as the language for this program as it has mature tools, including a debugger and profiler. The definitions of some important CUDA terms are given in Table \ref{tab:cuda}. I created the new program from scratch in CUDA, taking inspiration from in-house CPU code.  

To efficiently solve a problem with parallelism using CUDA on a GPU, the program should have the following qualities\cite{weigel2011simulating}:
\begin{enumerate}
\item Local calculations, keeping the need for communication between threads to a minimum.
\item Coherent threads with as little thread divergence as possible.
\item A number of threads much greater than available multiprocessors.
\item Many more arithmetic operations and use of shared memory than global memory accesses.\\
\end{enumerate}

\begin{figure}[!h]\centering
\subfloat[Time per fold against threads per block for the HP $6\times6$ model.]{\includegraphics[width=0.47\textwidth]{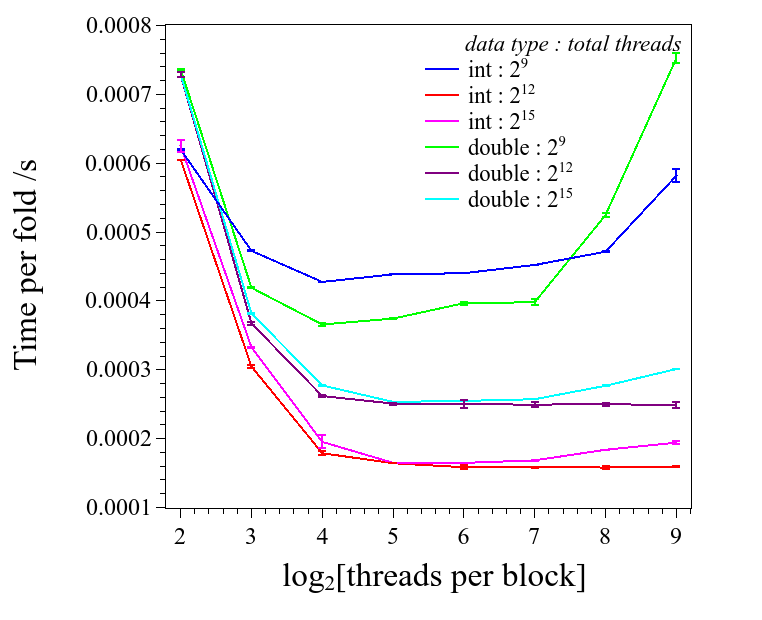}}\\
\subfloat[Time per fold against threads per block for the HP $3\times3\times3$ model.]
{\includegraphics[width=0.47\textwidth]{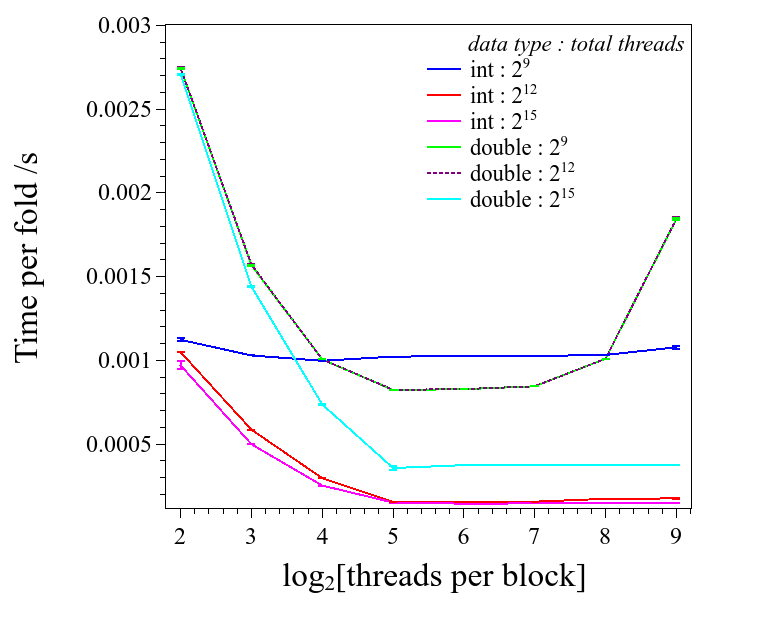}}
\caption{ Time per fold against threads per block for different numbers of total threads and for \textit{int} and \textit{double} variables. The three variables that effect the running time of the CUDA program are the data type used  (integer or double), the total number of threads launched by the kernel and the way these threads are split into blocks. The size of each block is important as each multiprocessor core runs 1 block at a time. These results show that the fastest set ups for the HP model on the $6\times6$ and $3\times3\times3$ lattices use integer variables, at least $2^{15}$ threads and threads per block values of 64 ($2^6$).}
\label{fig:gridblock}\end{figure}

To optimize the running time of code on the GPU, it is important to ensure that the threads per block and block sizes specified in the launch of a CUDA kernel do not increase latency. Threads are launched in warps of 32 and so to reduce any wasted processors, threads per block should be a multiple of 32. The grid size and total threads per block is best determined within the context of the model. I ran tests for varying block and grid size parameters to find the optimum set up.

I converted the floating point operations to integer operations in order to avoid issues with rounding floating point variables; if floating points are used, results can vary for different GPU architectures and between the GPU and the CPU\cite{fp_handbook}. Integer precision arithmetic operations sometimes use fewer cycles and therefore run faster than those using floats or doubles\cite{doublefloat}. In order to examine the effect on the speed of the code when using integer or double variables, I converted the decimal values in the energy matrices (Table \ref{tab:hp} and Appendix \ref{app:MJ}) to integer values and ran the code on the GeForce GTX 750Ti GPU. The speed of the code using integers and doubles is compared in Figure \ref{fig:gridblock} which shows that when comparing optimal integer and double runs, the integer computations run approximately twice as fast as those using doubles. Figure \ref{fig:gridblock} also shows the comparisons between different total thread numbers and block sizes. Code with over $2^{15}$ total threads runs approximately four times as fast as that with $2^{9}$. The optimal threads per block are shown by the minima of the graphs. The fastest configurations for the HP model on both the $6\times6$ and $3\times3\times3$ lattice used 64 threads per block. The future runs were executed with these optimized parameters. 

I wrote the core program for the project with the capability to run the HP or MJ models on either the $5\times5$, $6\times6$ or $3\times3\times3$ lattice. I verified the accuracy of the code by running cross checks against the results of Helling et al\cite{helling2001designability}. I verified that the code gave the same most designable folds for the MJ model on the $6\times6$ lattice and HP $3\times3\times3$ lattice as in Helling et al. (Figure \ref{fig:top_folds}). Other various cross-checks were carried out and can be seen in Appendix B.

\subsection{Designability, robustness and the topology of GP maps.}

The concepts of designability and robustness can be expressed mathematically and used to classify the topology of the GP map\cite{schaper2014arrival}. In this report the word `fold' is used interchangeably with `phenotype' and the word `sequence' is used interchangeably with `genotype'.  For a model with alphabet size $K$ and genotype length $L$ there are $K^L$ possible genotypes. The designability $N_q$ of a phenotype $q$ is defined as the size of the set of genotypes that have phenotype $q$.   
To define robustness we must talk about the neighborhood of a genotype and a quantity called $\phi_{pq}$. The `neighborhood' of a genotype is all other sequences that differ from it by a single point mutation. The number of neighbors of a sequence is therefore $(K-1)L$. If the designability (number of genotypes) of a phenotype is $N_{q}$ then the total number of neighbors of the phenotype is given by $N_{q}(K-1)L$. 

$\phi_{pq}$ is the average probability of a mutation from phenotype $q \Rightarrow p$. It is therefore the fraction of neighbors of $q$ that map to $p$;
\begin{equation}
\phi_{pq} = \frac{1}{N_{q}} \displaystyle \sum_{i=1}^{N_{q}} \Phi_p(g_i).
\end{equation}
%
%
where $\Phi_p(g_i)$ is the fraction of the $(K-1)L$ neighbors of genotype $g_i$ that result in phenotype $p$w and the sum averages over the $N_q$ genotypes that fold to phenotype $q$. 

\begin{figure}[h] \centering
\includegraphics[width=0.85\linewidth]{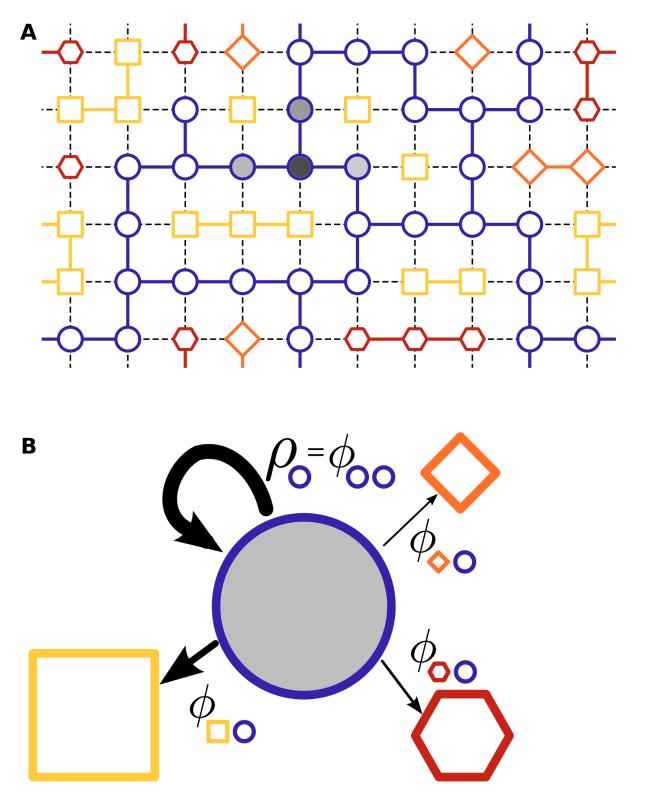}
\caption{(A) A hypothetical genotype space with different colors and shapes representing different phenotypes. Neutral connections between genotypes are shown in color, highlighting the neutral networks formed. (B) A depiction of how mutations of the original phenotype contribute to $\phi_{ij}$ values, the thickness of the arrows represent the size of the $\phi$ value. Here the robustness has the largest $\phi$ value and mutation from the circle phenotype to the orange diamond has the smallest $\phi$ value. Taken from\cite{schaper2014arrival}.}
\label{fig:sum44}
\end{figure}

The phenotype robustness can now be defined as $\rho_p = \phi_{pp}$. Robustness is the average probability of a neutral mutation $p \Rightarrow p$. Another interesting property of a phenotype is $\phi_{del}$ which measures the probability a mutation will result in the deleterious phenotype. The deleterious phenotype in the protein folding model is assigned to amino acids with no unique ground state fold. An example of a genotype space and the nature of different $\phi_{pq}$ values is shown in Figure \ref{fig:sum44}\cite{schaper2014arrival}.

Before discussing the expectations of these values in the `plum pudding' and `spaghetti' models I introduce the null model for quantitative comparison. The frequency of phenotypes is always a constraint defined by the physics of the problem (Equation \ref{eq:energy}). It is the location of phenotypes in the space that is variable. In the null model phenotypes are randomly located in the genotype space. The phenotypes of neighbors in the null model are therefore not correlated. A relevant quantity in the null model is the threshold $\gamma_{pq}$; this is the value of the global frequency of phenotype $p$ such that it will appear on average once in the neighborhood of $q$. Below this frequency the expected number of occurrences of $p$ in the neighborhood of $q$ drops below 1;
\begin{equation}
\gamma_{pq} = \frac{1}{N_{q}(K-1)L}.
\end{equation}

In the `plum pudding' model, the only significant values of $\phi$ are $\phi_{pp}$ (the robustness) and $\phi_{del}$ as the `plums' make limited connection with others. In the `spaghetti bowl' model, many values of $\phi_{pq}$ would be greater than the null model for many phenotypes $q$ as the `spaghetti' phenotypes make a greater number of contacts with other phenotypes. 

\begin{figure}[h]
\includegraphics[width=\linewidth]{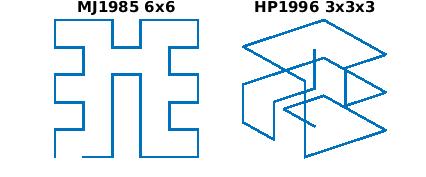}
\caption{The most designable fold for the MJ model on the $6\times6$ lattice and the HP model on the $3\times3\times3$ lattice.}
\label{fig:top_folds}
\end{figure}

All $2^{27} \approx 1.34 \times 10^{8}$ sequences in the HP $3\times3\times3$ model were exhaustively enumerated to produce and analyze the full GP map for the HP $3\times3\times3$ model. $\phi_{pq}$ was calculated for the neighborhood of the most designable fold (see Figure \ref{fig:top_folds}). The robustness and $\phi_{del}$ were calculated for every phenotype in the model.

The phenotypic robustness, $\phi_{pq}$ and $\phi_{del}$ in the $3\times3\times3$ HP model were compared to what would be expected from the `spaghetti', `plum pudding' and null models. 

\paragraph{Complexity and Designability}

Current investigations into the GP maps of RNA and gene networks have suggested connections between complexity and designability (Louis et al., unpublished). This suggestion prompted the investigation within this project into the relationship between the complexity and designability on the $6\times6$ lattice model. 

There are many definitions of complexity and the measurement method can vary between models. One method pioneered by Kolmogorov is called Kolmogorov complexity\cite{kolmogorov1965three}. The Kolmogorov complexity of a string $x$ is defined as `the size of the shortest string $y$ from which the universal Turing machine produces $x$'\cite{watanabe1992kolmogorov}. Exact Kolmogorov complexity is technically incomputable but there are many methods for approximating it. One method used in this project found an upper bound of the Kolmogorov complexity using a compression technique. The process is to write the fold as a sequence of Up, Down, Left and Right (numerically represented by 0,1,2,3) and then to compress this string and calculate its resultant length. This compression estimate of Kolmogorov complexity was compared to designability for the $6\times6$ phenotypes.

\begin{figure}[!h] \centering
\includegraphics[width=0.5\linewidth]{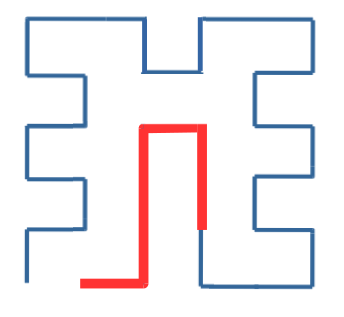}
\caption{The most designable fold for the MJ model on the $6\times6$ lattice. The minimum defining chain length for this fold given the lattice and start point is shown in red. Once the red line is fixed, the rest of the fold is completely determined.}
\label{fig:mdcl}
\end{figure}

Another approximation to the Kolmogorov complexity I used was the minimum defining chain length (MDCL). It has been suggested that the minimum length of unique chain in a fold correlates with its designability (Dingle, unpublished). The calculation of MDCL comprises finding the shortest path for each fold such that once that path is followed, given the starting position and size of the lattice, the rest of the fold is fixed. This is a promising approximation to the Kolmogorov complexity as it is analogous to the shortest definition of a fold. For example, the most designable fold for the the MJ $6\times6$ model has a short defining chain length of 7, shown by the red path in Figure \ref{fig:mdcl}. 

An attempt was made to calculate the MDCL for all compact phenotypes on the $6\times6$ lattice. These estimates of complexity for each fold were compared with designability for the folds to investigate whether there was any correlation between them.

%
%

\section{Results and Discussion}

I used CUDA to write programs for the GPU that calculated the ground state fold for every genotype for a given lattice and alphabet. The GPU gave a significant speed up which enabled various explorations of the properties of the genotype-phenotype spaces of various models.

%
%

\subsection{GPU Code Design and Acceleration}

The code was designed to fit the criteria in Section \ref{sec:code}. Each thread generated or imported a sequence and computed the sequence's lowest energy fold using arithmetic operations. The threads were completely independent and coherent in the operations they applied. The number of threads launched was of the order of $2^{20-28}$ and as such there were many more threads than processors. Finally, the number of global memory accesses were minimal as the majority of each threads calculations were either arithmetic operations or comparisons with values stored in registers. The CUDA code is in Appendix C and can be seen to exhibit all of these properties.

The speed ups shown in Table \ref{tab:speedup}  were obtained on the Tesla C2070 card, compiled with NVCC using CUDA v. 3.2\cite{nickolls2008scalable} and were compared to speeds on the Intel® Core™ i5-480M CPU.  The values in Table \ref{tab:speedup} were from the random sampling of some of possible sequences due to the impracticality of full enumeration on the CPU. Full enumeration is even faster per fold on the GPU as fewer memory operations have to be carried out. Full enumeration is faster because threads on the GPU do not have to access random numbers in global memory which have been transferred from the CPU.

\begin{table}[h]
\caption{Time per fold for CPU v GPU and the equivalent speed up per sequence.}
\centering
\begin{tabular}{rllr}

 Model & CPU /s & GPU /s & Speed up \\
\hline
MJ $3\times3\times3$ & 0.162  & 0.00028  & 580  \\
HP $3\times3\times3$ & 0.160 & 0.00026 & 620 \\
MJ $6\times6$ & 0.083 & 0.000119 & 700 \\
HP $6\times6$ & 0.081 & 0.000135 & 600 \\
\hline
\end{tabular}
\label{tab:speedup}
\end{table}

A speed up of 580-700 fold, as shown in Table \ref{tab:speedup}, is very large reducing sampling that would take days or years to minutes or days respectively. For example, the full enumeration of the HP $6\times6$ model, which would take over 180 years on a CPU is underway at time of writing and should take a total of approximately 110 days to complete on a single GPU chip. It is worth noting that the GPU provided approximately equal accelerations for calculations on the $6\times6$ and $3\times3\times3$ lattices. This is because the limiting factor on both the CPU and GPU was the number of folds that have to be tested for each sequence. This determined the number of calculations required on the CPU and determined the number of memory accesses on the GPU. There are twice as many folds on the $3\times3\times3$ lattice compared to the $6\times6$ lattice and as such the GPU and CPU ran approximately twice as fast on the $6\times6$ lattice compared to the $3\times3\times3$ lattice.

\subsection{Designability, robustness, \textbf{$\phi_{pq}$} and the topology of the HP $3\times3\times3$ GP map}  

All of the results in this subsection are for the HP model on the $3\times3\times3$ lattice. The \textbf{null model} used for comparison in these results is one in which the location of phenotypes within the genotype space is random. 

The full GP map for this model was enumerated. The distribution of designability for this model is shown in Appendix B as part of the code verification.
With the phenotype for all genotypes known, I calculated the robustness of all the model's phenotypes. I then examined the phenotypes in the neighborhood of the top fold $\phi_{pq}$, and the connections to the deleterious phenotype in the neighborhood of all phenotypes. I used these results to deduce how well the model is described by both the `spaghetti' and `plum pudding' metaphors. 

When all sequences in this model are enumerated, there are 4255 out of 51704 possible folds which do not occur at all as ground states in the  $3\times3\times3$ model, i.e. they have zero designability. This means that there are only 47449 possible phenotypes for sequences on this lattice. In this model, 4.76\% of all sequences have a ground state fold. All other genotypes have non-unique ground states and therefore 95.24\% of sequences have the deleterious phenotype.

%
%

\paragraph{Robustness}

\begin{figure}[h]
\includegraphics[width=\linewidth]{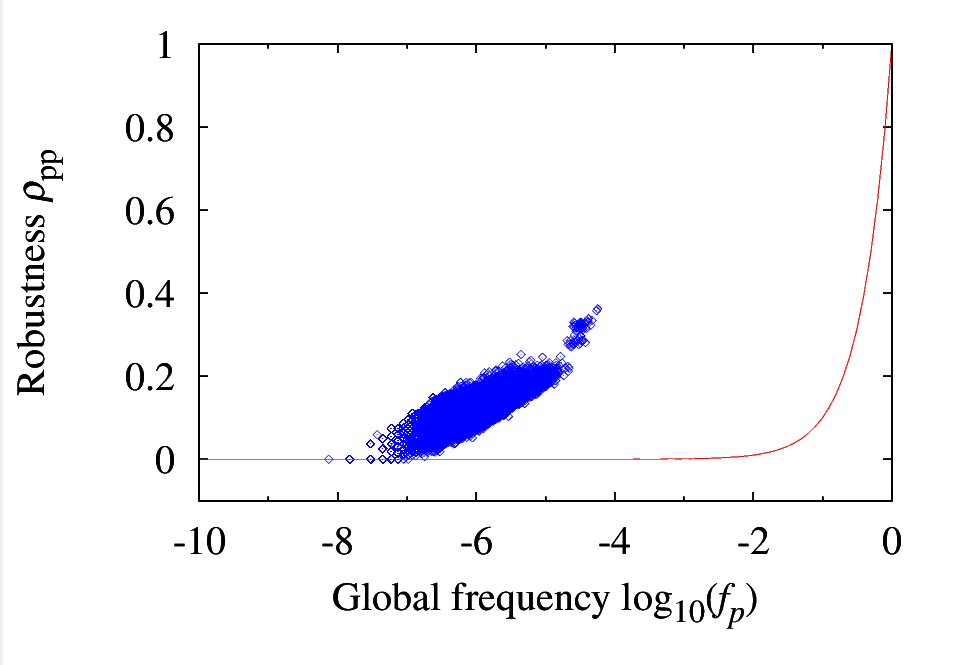}
\caption{Robustness against designability for all folds in the HP model on $3\times3\times3$ lattice. In the null model designability is equal to robustness (red line). The frequency of neutral mutations in the null model would fit this line as it would correlate exactly with the global frequency of that phenotype.}
\label{fig:3x3x3robustness}
\end{figure}

The robustness against designability for all phenotypes in the HP $3\times3\times3$ model is shown in Figure \ref{fig:3x3x3robustness}.

The robustness for a significant number of phenotypes (43.6\%) is larger than the null model. This suggests that in this HP model, many phenotypes are likely to have neutral networks and thereby percolate the space. The points that fall on the null model (red line) are phenotypes with robustness equal to what would be expected if there were no correlations between a fold and its neighbor. The large number of phenotypes with significantly larger robustness than the null model is evidence that a given phenotype is correlated with its neighborhood.

%
%

\paragraph{Phenotypes in the neighborhood of the top fold}
$\phi_{pq}$ is the frequency of mutations that result in phenotype $p$ in the single mutation neighborhood of phenotype $q$. Here I looked at the neighborhood of the most designable fold.
\begin{figure}[h]
\includegraphics[width=\linewidth]{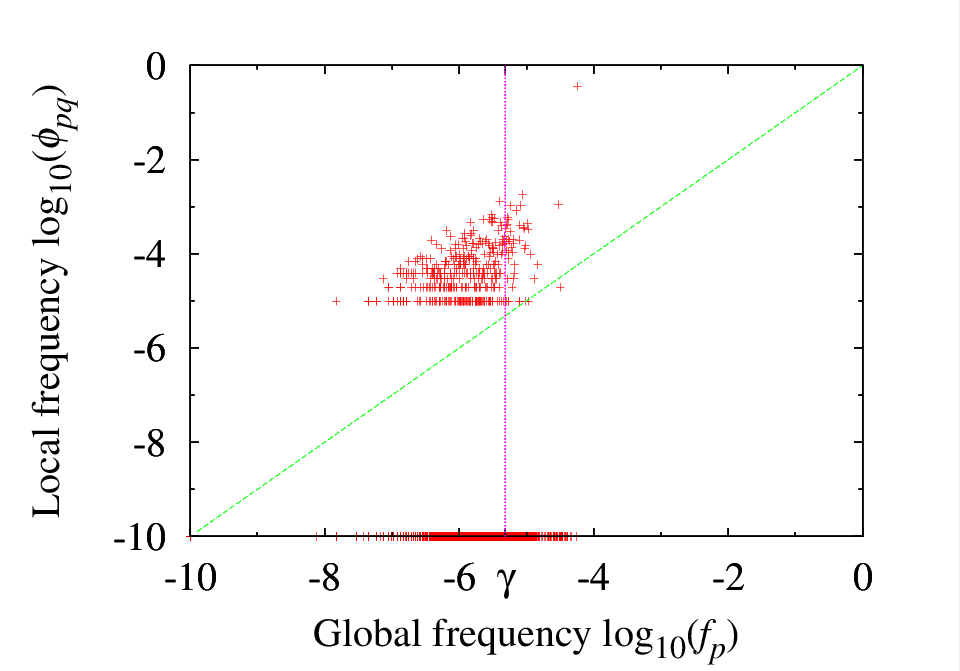}
\caption{Log $\phi_{pq}$ against log frequency for connections to the top fold. Many phenotypes, despite having a frequency below the threshold are connected to the top fold for the $3\times3\times3$. The green line shows the null model, for which the frequency of phenotypes in the neighborhood of the top fold is random and so is equal to their global frequency. The pink vertical line shows the threshold $\gamma$, below which the phenotype is so infrequent that it is not expected to occur in a randomly connected GP map. The top right point is the robustness $\phi_{pp}$ (connections from the most designable fold back to itself). Due to the need for a log scale, phenotypes with a zero frequency are represented with a frequency of $10^{-10}$ at the bottom of the graph.}
\label{fig:phiHP19963x3x3}
\end{figure}
Figure \ref{fig:phiHP19963x3x3} shows the local frequency of phenotypes in the neighborhood of the most designable fold against their global frequency. The distribution that would be produced by the null model is shown by the green line in Figure \ref{fig:phiHP19963x3x3}. The graph shows that in the HP $3\times3\times3$ model there are many phenotypes which have more connections to the most designable phenotype than would be expected in the null model (i.e. those above the green diagonal). The graph also shows that below the threshold $\gamma$ there are many connections from the most designable fold to other phenotypes; the existence of these connections would not be expected in the null model. Those phenotypes with a zero value ($10^{-10}$) of local frequency do not occur in the immediate neighborhood of this phenotype. The robustness of the top fold is the largest $\phi_{pq}$ value at the top right of the graph. This graph shows that due to its large robustness, the most designable fold is likely to percolate the space and form a large neutral network. The neighborhood of a phenotype is not random and is correlated with the phenotype itself. The large number of non-zero connections to other phenotypes from the top fold's phenotype suggests that the HP $3\times3\times3$ model is not `plum pudding'. In the HP $3\times3\times3$ model, 0.86\% of possible phenotypes are connected to the top fold, a larger value than would be predicted in the `plum pudding' model.
%
%
\paragraph{Deleterious phenotypes}
The distribution of connections to the deleterious phenotype is shown in Figure \ref{fig:phidel3x3x3}. In the null model the local frequency of deleterious mutations would be simply equal to the global frequency of deleterious mutations which is 0.9524. The figure shows that the deleterious phenotypes are underrepresented compared to the null model with a mean of 0.8618.
\begin{figure}[h]
\includegraphics[width=\linewidth]{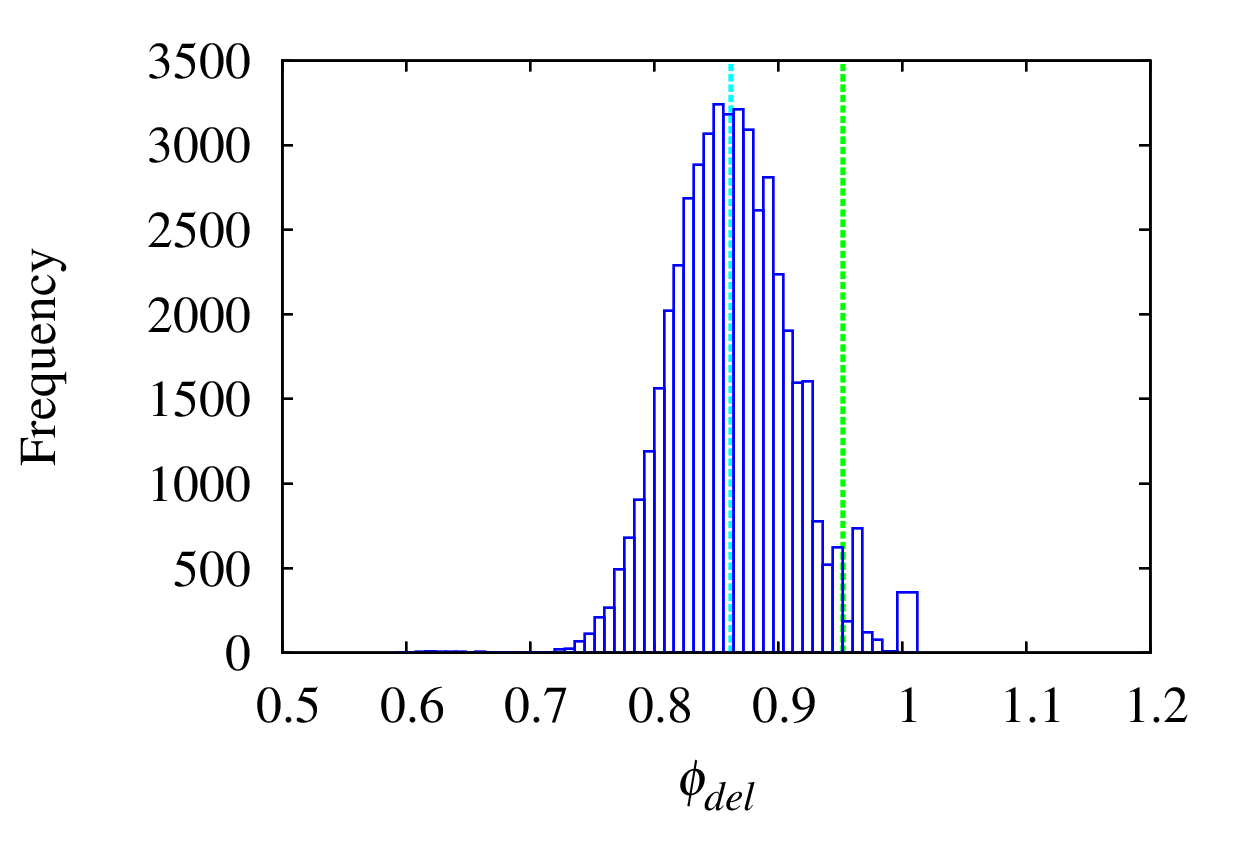}
\caption{Local frequency of deleterious phenotypes $\phi_{del}$ compared to global frequency of deleterious phenotypes $f_{del}$ for all phenotypes (folds) in the HP model on the $3\times3\times3$ lattice. The mean value of the distribution is marked in cyan. The randomised null model would have a distribution centered around $f_{del}=0.9524$, shown in green.}
\label{fig:phidel3x3x3} 
\end{figure}
Taking the robustness and $\phi_{del}$ together I investigated whether all of the non-deleterious neighbors were accounted for by the robustness of a phenotype. A histogram of the fraction of neighbors that are neither deleterious nor neutral is shown in Figure \ref{fig:nonneu}. I found that only 890 phenotypes (1.9\%) are completely isolated from any other and on average 4.0\% of a phenotypes neighbors are other, non-deleterious phenotypes. This means that in this HP model, the vast majority of the phenotypes possess many more connections to other phenotypes than if the model was described by the `plum pudding'. The main contribution to the histogram for the `plum pudding' model would be from phenotypes with zero distinct neighbors. Figure \ref{fig:nonneu} therefore presents compelling evidence that the HP model on the $3\times3\times3$ lattice is not well represented by the `plum pudding' model; the phenotypes of the GP map for the HP $3\times3\times3$ model are better described by the `spaghetti bowl' metaphor.

\begin{figure}[h]
\includegraphics[width=0.9\linewidth]{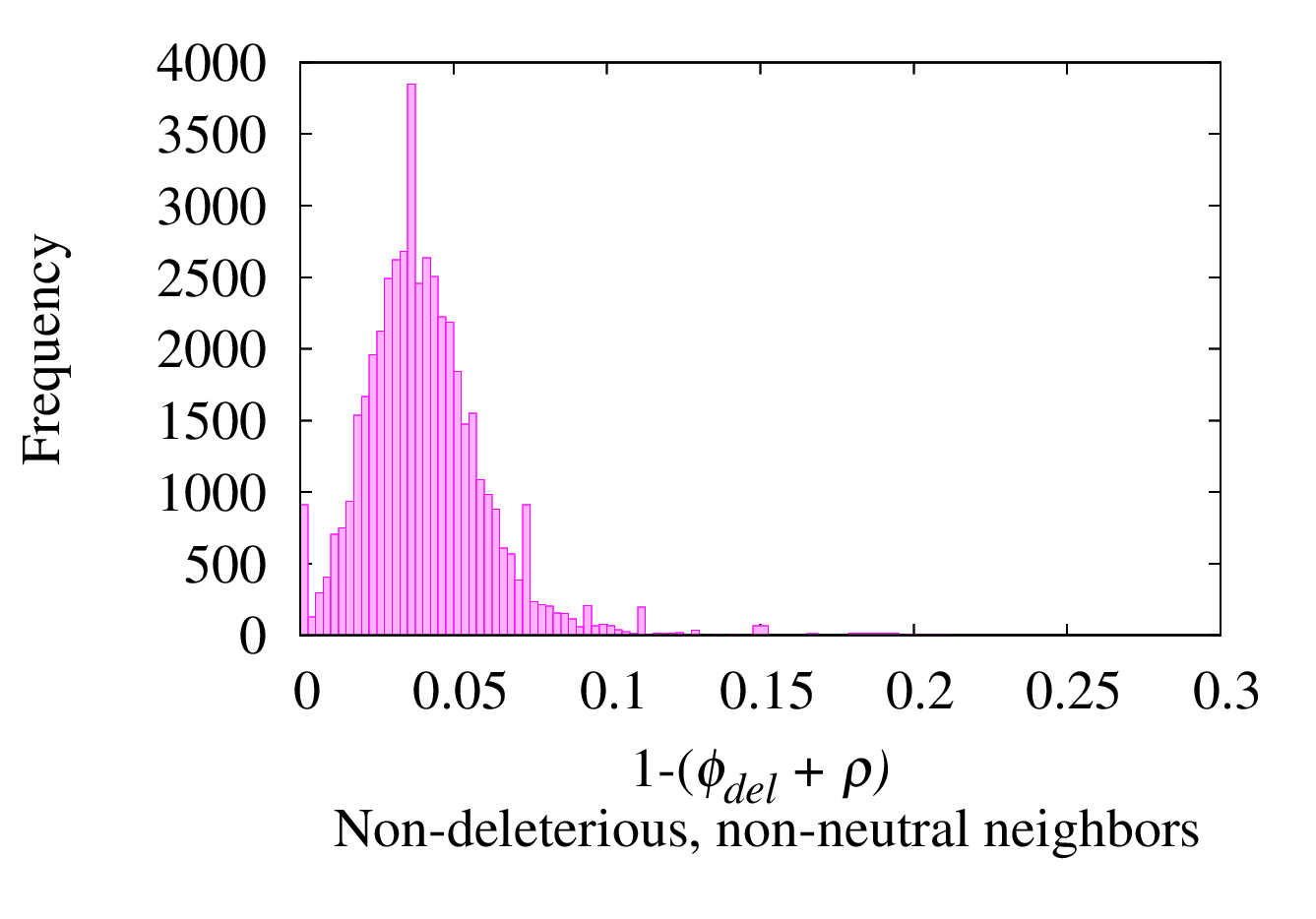}
\caption{A histogram of the frequency of non-deleterious, distinct phenotypes $\phi_{del}$ in the neighborhood of all phenotypes in the HP model on the $3\times3\times3$ lattice. On average 4.0\% of neighbors of a phenotype are non-deleterious, non-neutral neighbors.}
\label{fig:nonneu} 
\end{figure}

The data shown in Figures \ref{fig:3x3x3robustness}-\ref{fig:phidel3x3x3} shows that the HP model on the $3\times3\times3$ lattice fits neither the `plum pudding' nor `spaghetti bowl' metaphors perfectly. The GP map contains at least one large and well connected neutral network (Figure \ref{fig:phiHP19963x3x3}) as well as many other significant neutral networks (Figure \ref{fig:3x3x3robustness}). However, 98.1\% of phenotypes are connected to other phenotypes (Figure \ref{fig:nonneu}) meaning this GP map is best described by the `spaghetti' model. The distinction cannot be made between all phenotypes being connected via mutation or forming a small number of subnetworks. The latter can be thought of as balls of `spaghetti'.
 
%
%
 
\subsection{Complexity and Designability}

The complexity-designability relationship can be explored by sampling the genotype space because a sufficient distribution of the most common folds (designability) occurs without full enumeration. 
Only the $6\times6$ model was used to fully analyze relationships between complexity and designability because spotting visual distinctions between folds is easier on the 2D lattice. The 50 most designable and 50 least designable structures are shown in Appendix B. Visual inspection shows that they are clearly different. The challenge is to discover whether this difference is correlated with a simple complexity measure.

\begin{figure}[h]
\includegraphics[width=\linewidth]{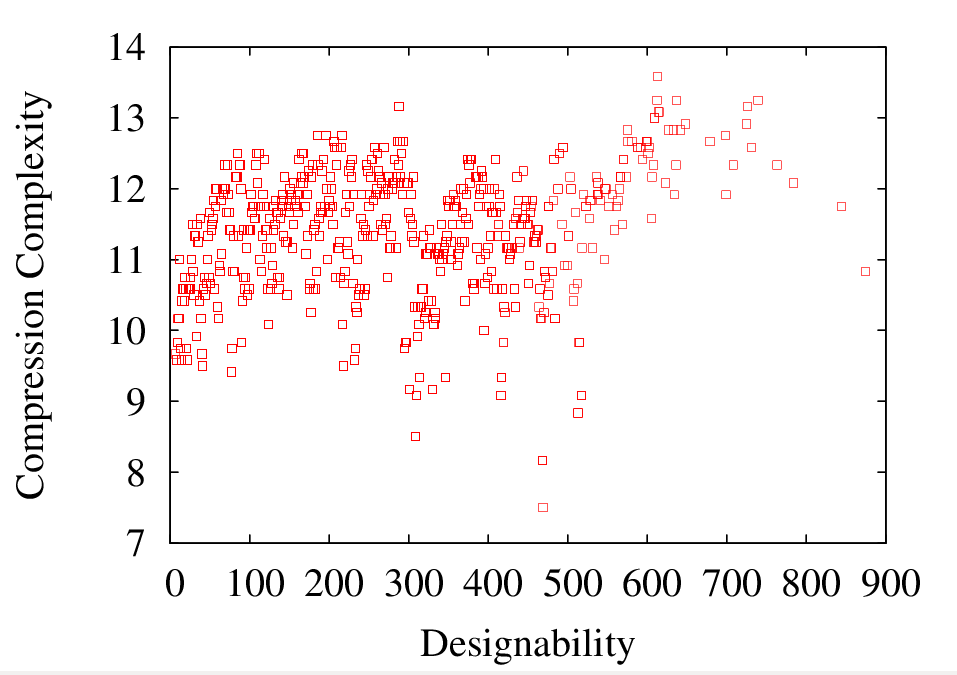}
\caption{Compression complexity against designability for the MJ model on the $6\times6$ lattice (for $2^{25}$ random sequences). There is no correlation between complexity and designability for this measure. }
\label{fig:compress}
\end{figure}

Figure \ref{fig:compress} shows the compression complexity approximation against designability, there is no pattern that emerges from this comparison.

\begin{figure}[h]
\includegraphics[width=\linewidth]{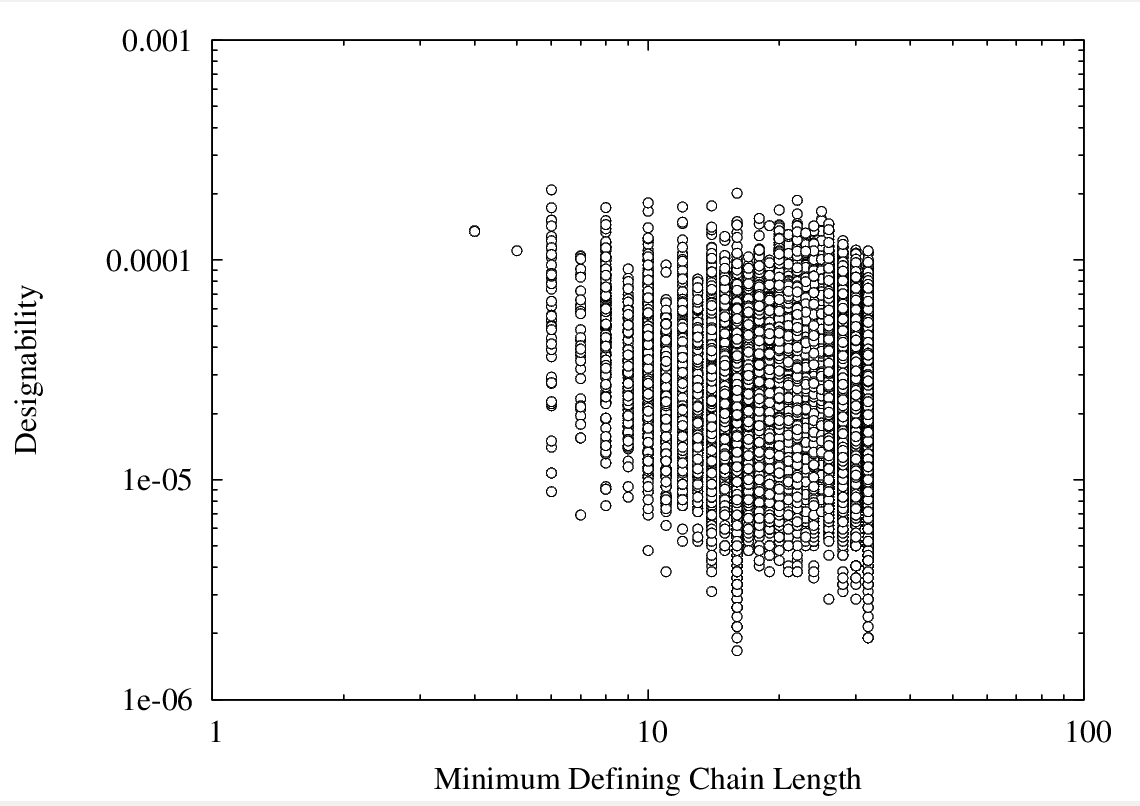}
\caption{Minimum defining chain length against designability for the MJ model on the $6\times6$ lattice (for $2^{25}$ random sequences). There is no clear correlation between MDCL and designability for this model. }
\label{fig:min_des19}

\end{figure}

Figure \ref{fig:min_des19} shows the minimum defining chain length (MDCL) against designability. There is no clear correlation but there is some indication that the least designable folds have long minimum defining chain lengths. It is possible that with improved calculation of the MDCL a clearer correlation may emerge, as the current measure does not recognize the MDCL correctly for all folds. This method of complexity measure deserves further consideration because it is clear that the MDCL is a form of minimum information content. This could be biologically relevant as it suggests that short motifs may define larger structural features that are not necessarily conserved in their amino acid sequence. The $6\times6$ lattice is, of course, artificial in its constraints and as such, any future connection found must still be interpreted within that context.

%
%

\newpage
\section{Conclusions}
This particular GP mapping problem was extremely well suited to the application of GPU computing. The very large speed up (580-700 fold) was due to the properties of the HP and MJ models matching many of those required for an efficient GPU application. These qualities are typical of GP mapping: locality, coherence, the dominance of arithmetic calculations and multitude of calculations needed. This means many GP mappings are likely to be greatly sped up on the GPU and the result can therefore be generalised. This applicability is already evident in my work with Chico Camargo (University of Oxford) who has produced a 1000-fold speed up in the calculation of the GP map for gene networks by implementing his model on a GPU, for which he used concepts from my code (Appendix C).

There are two simple improvements that could be made to further accelerate the time per fold I have achieved. Firstly, a network of GPUs would increase computational throughput. Secondly, the generation of random numbers on the GPU would reduce the number of required memory accesses. This is in comparison to the current method which requires copying large quantities of random numbers from the CPU. This would increase performance for any model requiring random sampling. Due to the relevant in-expense of GPUs and the ever expanding range of tool-kits provided by NVIDIA, these options would both be worth pursuing in further research. A full enumeration of the HP model on the $6\times6$ lattice on the GPU is currently running and will allow the exploration of its full GP map. This was previously impossible due to the extreme length of time it would have taken (approximately 180 years) before this acceleration. 

The investigation of the $3\times3\times3$ HP model has shown that contrary to Goldstein's postulation\cite{Goldstein2008}, the HP model does not fit the description of `plum pudding' like and that their results are an artefact of the small size of their lattice. It is important to bear in mind that the HP model on the $3\times3\times3$ lattice has a very large genotype space ($2^{27}\approx1.3\times10^8$ genotypes) and 95.24\% of its genotypes have the deleterious phenotype (do not fold). If the GP map, given these constraints, fitted the `plum pudding' model, we would expect to see more than 1.8\% of phenotypes being completely isolated and we would expect many fewer than 4\% of mutations to result in new phenotypes. This HP model is even more poorly described by the null model as there are significant correlations between neighbors. The HP model is best described by the `spaghetti' model but no simplistic metaphor describes this space perfectly.
Future work should be done on GP maps for both the MJ model and larger lattices in order to investigate how the topology changes with larger alphabets and lattices. The MJ model would also allow the  investigation of how the connections and neutral networks within a GP map are effected by changes to the energy threshold.

The attempt to discover underlying properties of phenotypes that may correlate with designability on the $6\times6$ lattice did not reveal any conclusive results. However, further investigation into the relationship between complexity and designability would be worthwhile. In particular, the computation of defining chain length can be perfected as the current complexity measure is incomplete.

This project has obtained a very significant acceleration using GPU methods which is highly translatable to both other attempts at GP mapping and NP complete problems. I have been able to analyze the topology of a new GP map and to enumerate the previously unobtainable 6$\times$6 GP map for investigation. In a wider context, these GP maps can provide insight into why there are relatively very few natural folds compared to what one might expect theoretically. If the defining properties of these structures could be found, they could be used predict novel stable folds, an area of great interest to the pharmaceutical industry\cite{floudas2006advances}. This project also provides the first demonstration of GPU acceleration for a lattice protein folding model and as such demonstrates the potential power of the application of parallel programming techniques to other biophysical models. 

\phantomsection
\section*{Acknowledgments} 

I would like to thank Professor Ard Louis, Chico Camargo, Dr Flavio Romano and Dr Sam Greenbury for their help in completing this project. 

%
%

\phantomsection
\bibliographystyle{unsrt}
\bibliography{thesis}

\begin{thebibliography}{10}

\bibitem{li1996emergence}
Hao Li, Robert Helling, Chao Tang, and Ned Wingreen.
\newblock Emergence of preferred structures in a simple model of protein
  folding.
\newblock {\em Science}, 273(5275):666--669, 1996.

\bibitem{de1900spaltungsgesetz}
Hugo De~Vries.
\newblock {\em Das Spaltungsgesetz der bastarde}.
\newblock Borntraeger, 1900.

\bibitem{mendel1866versuche}
Gregor Mendel.
\newblock Versuche {\"u}ber pflanzenhybriden.
\newblock {\em Verhandlungen des naturforschenden Vereines in Brunn 4: 3}, 44,
  1866.

\bibitem{morgan1919physical}
Thomas~Hunt Morgan.
\newblock {\em The physical basis of heredity}.
\newblock JB Lippincott, 1919.

\bibitem{fisher1919xv}
Ronald~A Fisher.
\newblock Xv.—the correlation between relatives on the supposition of
  mendelian inheritance.
\newblock {\em Transactions of the royal society of Edinburgh},
  52(02):399--433, 1919.

\bibitem{geer1962genotype}
BW~Geer and MM~Green.
\newblock Genotype, phenotype and mating behavior of drosophila melanogaster.
\newblock {\em American Naturalist}, pages 175--181, 1962.

\bibitem{schwartz2001genotype}
Peter~J Schwartz, Silvia~G Priori, Carla Spazzolini, Arthur~J Moss, G~Michael
  Vincent, Carlo Napolitano, Isabelle Denjoy, Pascale Guicheney, G{\"u}nter
  Breithardt, Mark~T Keating, et~al.
\newblock Genotype-phenotype correlation in the long-qt syndrome gene-specific
  triggers for life-threatening arrhythmias.
\newblock {\em Circulation}, 103(1):89--95, 2001.

\bibitem{johannsen2014genotype}
Wilhelm Johannsen.
\newblock The genotype conception of heredity.
\newblock {\em International journal of epidemiology}, 43(4):989--1000, 2014.

\bibitem{ciliberti2007innovation}
Stefano Ciliberti, Olivier~C Martin, and Andreas Wagner.
\newblock Innovation and robustness in complex regulatory gene networks.
\newblock {\em Proceedings of the National Academy of Sciences},
  104(34):13591--13596, 2007.

\bibitem{alberch1991genes}
P~Alberch.
\newblock From genes to phenotype: dynamical systems and evolvability.
\newblock {\em Genetica}, 84(1):5--11, 1991.

\bibitem{duret2008neutral}
L~Duret.
\newblock Neutral theory: the null hypothesis of molecular evolution.
\newblock {\em Nature Education}, 1(1):218, 2008.

\bibitem{masel2010gpmap}
Joanna Masel and Meredith~V. Trotter.
\newblock Robustness and evolvability.
\newblock {\em Trends in Genetics}, 26(9):406 -- 414, 2010.

\bibitem{lipman1991modelling}
David~J Lipman and W~John Wilbur.
\newblock Modelling neutral and selective evolution of protein folding.
\newblock {\em Proceedings of the Royal Society of London. Series B: Biological
  Sciences}, 245(1312):7--11, 1991.

\bibitem{izquierdo2008evolution}
Eduardo Izquierdo and Chrisantha Fernando.
\newblock The evolution of evolvability in gene transcription networks.
\newblock In {\em ALIFE}, pages 265--273. Citeseer, 2008.

\bibitem{wagner1996perspective}
Gunter~P Wagner and Lee Altenberg.
\newblock Perspective: complex adaptations and the evolution of evolvability.
\newblock {\em Evolution}, pages 967--976, 1996.

\bibitem{shackleton2000investigation}
Mark Shackleton, R~Shipma, and Marc Ebner.
\newblock An investigation of redundant genotype-phenotype mappings and their
  role in evolutionary search.
\newblock In {\em Evolutionary Computation, 2000. Proceedings of the 2000
  Congress on}, volume~1, pages 493--500. IEEE, 2000.

\bibitem{pigliucci2010genotype}
Massimo Pigliucci.
\newblock Genotype--phenotype mapping and the end of the ‘genes as
  blueprint’metaphor.
\newblock {\em Philosophical Transactions of the Royal Society B: Biological
  Sciences}, 365(1540):557--566, 2010.

\bibitem{smith1970natural}
JM~Smith.
\newblock Natural selection and the concept of a protein space.
\newblock {\em Nature}, 225(5232):563, 1970.

\bibitem{van1999neutral}
Erik Van~Nimwegen, James~P Crutchfield, and Martijn Huynen.
\newblock Neutral evolution of mutational robustness.
\newblock {\em Proceedings of the National Academy of Sciences},
  96(17):9716--9720, 1999.

\bibitem{rendel2011adaptive}
Mark~D Rendel.
\newblock Adaptive evolutionary walks require neutral intermediates in rna
  fitness landscapes.
\newblock {\em Theoretical population biology}, 79(1):12--18, 2011.

\bibitem{schaper2014arrival}
Steffen Schaper and Ard~A Louis.
\newblock The arrival of the frequent: How bias in genotype-phenotype maps can
  steer populations to local optima.
\newblock {\em PloS one}, 9(2):e86635, 2014.

\bibitem{ferrada2012comparison}
Evandro Ferrada and Andreas Wagner.
\newblock A comparison of genotype-phenotype maps for rna and proteins.
\newblock {\em Biophysical journal}, 102(8):1916--1925, 2012.

\bibitem{kolodny2013universe}
Rachel Kolodny, Leonid Pereyaslavets, Abraham~O Samson, and Michael Levitt.
\newblock On the universe of protein folds.
\newblock {\em Annual review of biophysics}, 42:559--582, 2013.

\bibitem{nochomovitz2006highly}
Yigal~D Nochomovitz and Hao Li.
\newblock Highly designable phenotypes and mutational buffers emerge from a
  systematic mapping between network topology and dynamic output.
\newblock {\em Proceedings of the National Academy of Sciences of the United
  States of America}, 103(11):4180--4185, 2006.

\bibitem{wang2000symmetry}
Tairan Wang, Jonathan Miller, Ned~S. Wingreen, Chao Tang, and Ken~A. Dill.
\newblock Symmetry and designability for lattice protein models.
\newblock {\em The Journal of Chemical Physics}, 113(18):8329--8336, 2000.

\bibitem{li2009introduction}
Ming Li and Paul~MB Vit{\'a}nyi.
\newblock {\em An introduction to Kolmogorov complexity and its applications}.
\newblock Springer Science \& Business Media, 2009.

\bibitem{bonchev1976symmetry}
D~Bonchev, D~Kamenski, and V~Kamenska.
\newblock Symmetry and information content of chemical structures.
\newblock {\em Bulletin of Mathematical Biology}, 38(2):119--133, 1976.

\bibitem{anfinsen1973principles}
Christian~B Anfinsen et~al.
\newblock Principles that govern the folding of protein chains.
\newblock {\em Science}, 181(4096):223--230, 1973.

\bibitem{kallberg2012template}
Morten K{\"a}llberg, Haipeng Wang, Sheng Wang, Jian Peng, Zhiyong Wang, Hui Lu,
  and Jinbo Xu.
\newblock Template-based protein structure modeling using the raptorx web
  server.
\newblock {\em Nature protocols}, 7(8):1511--1522, 2012.

\bibitem{oldziej2005physics}
S~O{\l}dziej, C~Czaplewski, A~Liwo, M~Chinchio, M~Nanias, JA~Vila, M~Khalili,
  YA~Arnautova, A~Jagielska, M~others Makowski, et~al.
\newblock Physics-based protein-structure prediction using a hierarchical
  protocol based on the unres force field: assessment in two blind tests.
\newblock {\em Proceedings of the National Academy of Sciences of the United
  States of America}, 102(21):7547--7552, 2005.

\bibitem{wu2007ab}
Sitao Wu, Jeffrey Skolnick, and Yang Zhang.
\newblock Ab initio modeling of small proteins by iterative tasser simulations.
\newblock {\em BMC biology}, 5(1):17, 2007.

\bibitem{dill1985theory}
Ken~A Dill.
\newblock Theory for the folding and stability of globular proteins.
\newblock {\em Biochemistry}, 24(6):1501--1509, 1985.

\bibitem{miyazawa1985estimation}
Sanzo Miyazawa and Robert~L Jernigan.
\newblock Estimation of effective interresidue contact energies from protein
  crystal structures: quasi-chemical approximation.
\newblock {\em Macromolecules}, 18(3):534--552, 1985.

\bibitem{berger1998protein}
Bonnie Berger and Tom Leighton.
\newblock Protein folding in the hydrophobic-hydrophilic (hp) model is
  np-complete.
\newblock {\em Journal of Computational Biology}, 5(1):27--40, 1998.

\bibitem{mann2014exact}
Martin Mann and Rolf Backofen.
\newblock Exact methods for lattice protein models.
\newblock {\em Bio-Algorithms and Med-Systems}, 10(4):213--225, 2014.

\bibitem{berlekamp1978inherent}
Elwyn~R Berlekamp, Robert~J McEliece, and Henk~CA Van~Tilborg.
\newblock On the inherent intractability of certain coding problems.
\newblock {\em IEEE Transactions on Information Theory}, 24(3):384--386, 1978.

\bibitem{Goldstein2008}
Richard~A Goldstein.
\newblock The structure of protein evolution and the evolution of protein
  structure.
\newblock {\em Current Opinion in Structural Biology}, 18(2):170 -- 177, 2008.
\newblock Theory and simulation / Macromolecular assemblages.

\bibitem{luebke2004general}
David Luebke et~al.
\newblock General-purpose computation on graphics hardware.
\newblock In {\em Workshop, SIGGRAPH}, 2004.

\bibitem{owens2008gpu}
John~D Owens, Mike Houston, David Luebke, Simon Green, John~E Stone, and
  James~C Phillips.
\newblock Gpu computing.
\newblock {\em Proceedings of the IEEE}, 96(5):879--899, 2008.

\bibitem{weigel2011simulating}
Martin Weigel.
\newblock Simulating spin models on gpu.
\newblock {\em Computer Physics Communications}, 182(9):1833--1836, 2011.

\bibitem{shimokawabe201080}
Takashi Shimokawabe, Takayuki Aoki, Chiashi Muroi, Junichi Ishida, Kohei
  Kawano, Toshio Endo, Akira Nukada, Naoya Maruyama, and Satoshi Matsuoka.
\newblock An 80-fold speedup, 15.0 tflops full gpu acceleration of
  non-hydrostatic weather model asuca production code.
\newblock In {\em High Performance Computing, Networking, Storage and Analysis
  (SC), 2010 International Conference for}, pages 1--11. IEEE, 2010.

\bibitem{beberg2009folding}
Adam~L Beberg, Daniel~L Ensign, Guha Jayachandran, Siraj Khaliq, and Vijay~S
  Pande.
\newblock Folding@ home: Lessons from eight years of volunteer distributed
  computing.
\newblock In {\em Parallel \& Distributed Processing, 2009. IPDPS 2009. IEEE
  International Symposium on}, pages 1--8. IEEE, 2009.

\bibitem{creighton1990protein}
Thomas~E Creighton.
\newblock Protein folding.
\newblock {\em Biochemical journal}, 270(1):1--2, 1990.

\bibitem{brodtkorb2013graphics}
Andr{\'e}~R Brodtkorb, Trond~R Hagen, and Martin~L S{\ae}tra.
\newblock Graphics processing unit (gpu) programming strategies and trends in
  gpu computing.
\newblock {\em Journal of Parallel and Distributed Computing}, 73(1):4--13,
  2013.

\bibitem{10.1109/CGO.2007.13}
Ian Buck.
\newblock Gpu computing: Programming a massively parallel processor.
\newblock {\em Proceedings of the 2013 IEEE/ACM International Symposium on Code
  Generation and Optimization (CGO)}, 0:17, 2007.

\bibitem{nickolls2008scalable}
John Nickolls, Ian Buck, Michael Garland, and Kevin Skadron.
\newblock Scalable parallel programming with cuda.
\newblock {\em Queue}, 6(2):40--53, 2008.

\bibitem{fp_handbook}
J.M. Muller, N.~Brisebarre, F.~de~Dinechin, C.P. Jeannerod, V.~Lef{\`e}vre,
  G.~Melquiond, N.~Revol, D.~Stehl{\'e}, and S.~Torres.
\newblock {\em Handbook of Floating-Point Arithmetic}, pages 206--207.
\newblock Birkh{\"a}user Boston, 2009.

\bibitem{doublefloat}
A.~Arora and S.~Bansal.
\newblock {\em Unix and C Programming}, page 325.
\newblock Laxmi Publications Pvt Limited, 2005.

\bibitem{helling2001designability}
Robert Helling, Hao Li, R{\'e}gis M{\'e}lin, Jonathan Miller, Ned Wingreen,
  Chen Zeng, and Chao Tang.
\newblock The designability of protein structures.
\newblock {\em Journal of Molecular Graphics and Modelling}, 19(1):157--167,
  2001.

\bibitem{kolmogorov1965three}
Andrei~N Kolmogorov.
\newblock Three approaches to the quantitative definition ofinformation'.
\newblock {\em Problems of information transmission}, 1(1):1--7, 1965.

\bibitem{watanabe1992kolmogorov}
Osamu Watanabe.
\newblock {\em Kolmogorov complexity and computational complexity}.
\newblock Springer, 1992.

\bibitem{floudas2006advances}
CA~Floudas, HK~Fung, SR~McAllister, M~M{\"o}nnigmann, and R~Rajgaria.
\newblock Advances in protein structure prediction and de novo protein design:
  A review.
\newblock {\em Chemical Engineering Science}, 61(3):966--988, 2006.

\end{thebibliography}


\titleformat{\section}
  {\color{black}\large\sffamily\bfseries}
  {}
  {0em}
  {\colorbox{black}{\parbox{\dimexpr\linewidth-2\fboxsep\relax}{\appendixname~\thesection .}}}
  []

\newpage
\onecolumn

%
%

\begin{appendices}
\section{Result}
\subsection{Miyazawa-Jernigan contact energies}
\label{app:MJ}
\begin{figure*}[h!]
\includegraphics[width=\linewidth]{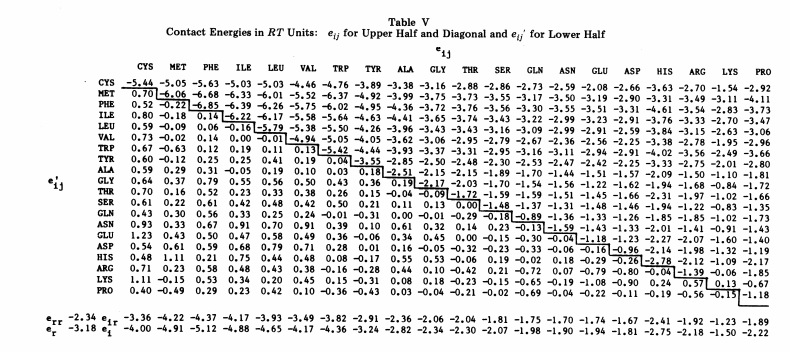}\\
\caption{The values including and above the diagonal of this matrix give the amino acid interaction energies\cite{miyazawa1985estimation}.}
\label{fig:MJmatrix}
\end{figure*}

The contact energies for the MJ1985 model are for the standard alphabet of 20 amino acids. The values used are from the work of Miyazawa and Jernigan\cite{miyazawa1985estimation} and are given in the matrix in Figure \ref{fig:MJmatrix}. 

\subsection{Checking the results of the GPU version of the model}
\begin{figure}[h]
\centering
\includegraphics[width=0.7\linewidth]{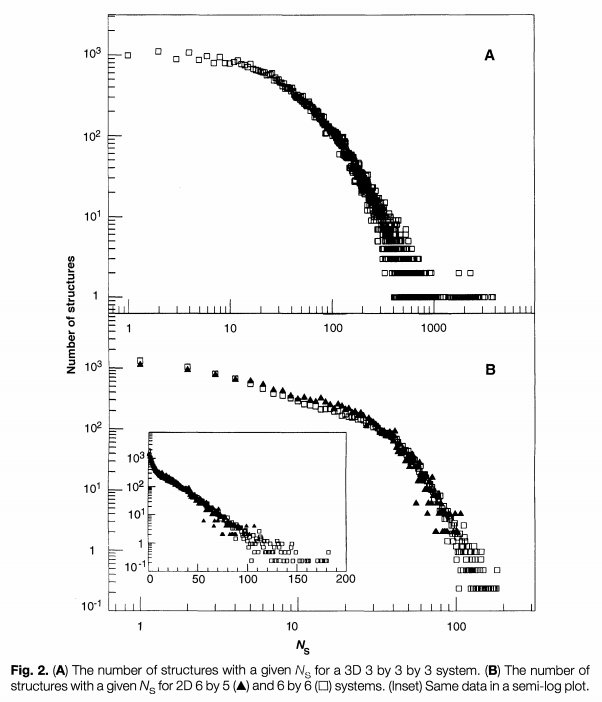}\\
\caption{Figure used for verification of the code from Li et al.\cite{li1996emergence}.}
\label{fig:li1996emergence}
\end{figure}

Figure \ref{fig:li1996emergence} is from Li et al. and along with the fold percentages and top fold data in Table \ref{tab:compare} was used to verify the results from the GPU program.

\begin{figure}[h!]
\centering
\includegraphics[width=0.6\linewidth]{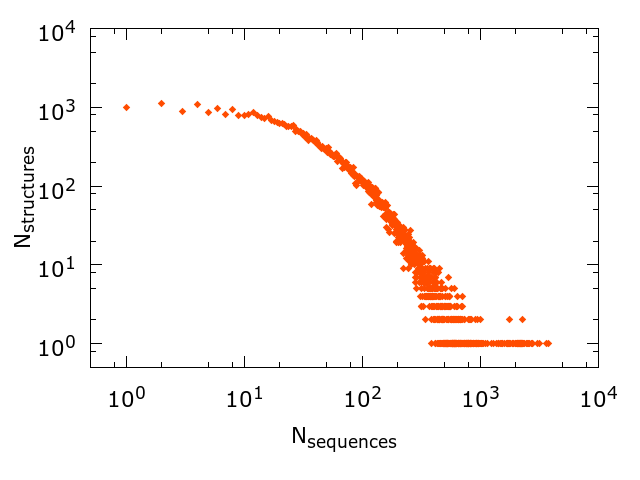}\\
\caption{Histogram of designability for the HP1996 model on the 3x3x3 lattice.}
\label{fig:HH_fwd_HP1996_3x3x3}
\end{figure}

The histogram for the full enumeration of the HP1996 3x3x3 lattice model is shown in Figure \ref{fig:li1996emergence}A. The data from this project (Figure \ref{fig:HH_fwd_HP1996_3x3x3}) is in agreement with Li et al. Note that only half the sequence-structure relationships have actually been computed in the process of forming these histograms because the GP map is completely symmetric.

\begin{figure}[h!]
\centering
\includegraphics[width=0.6\linewidth]{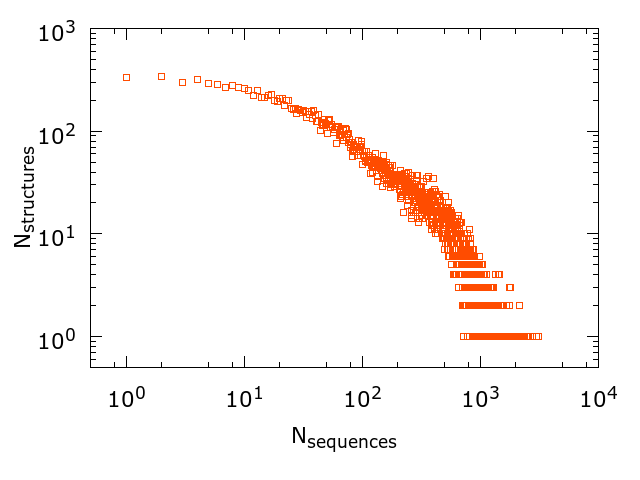}\\
\caption{Histogram of designability for the HP1996 model on the 6x6 lattice.}
\label{fig:HH_6x6}
\end{figure}

Figure \ref{fig:HH_6x6} shows the histogram of designabilities for the a random sampling of the HP1996 model on the 6x6 lattice. The top fold given in 1996 by Li et al.\cite{li1996emergence} is not identical to the fold we found. However, Li et al. used a comparably small sample, with the Li et al.'s most designable fold having only 200 sequences folding to it (Figure \ref{fig:li1996emergence}B) in contrast with the most designable fold in this project having over 3000 sequences folding to it (Figure \ref{fig:HH_6x6}). Their small sample size could mean the top ranking of their most designable fold is an artifact of the small sample. Furthermore, there is a highly unusual detail in Li et al.'s data (Figure \ref{fig:li1996emergence}B(inset)). The furthest right point is the one that has the highest designability and their graph suggests that there are in fact two folds that give this value, there is no mention of the second fold in the results and this strange distribution is probably also
a result of their small sample size.

\begin{table}[h]
\caption{Values from the GPU code compared with those of Li et al.\cite{li1996emergence}}
\centering
\begin{tabular}{rrrr}
\hline
 & & Li et al. & Project \\
\hline
\multirow{2}{*}{HP1996 $3\times3\times3$}
& Fold percentage & 4.75\%  & 4.76\%  \\
& $N_{s}$ top fold & 3794 & 3794 \\
\hline
\end{tabular}
\label{tab:compare}
\end{table}

\clearpage
\section{Designability and Complexity}
\begin{figure}[h!]
\centering
\includegraphics[width=0.8\linewidth]{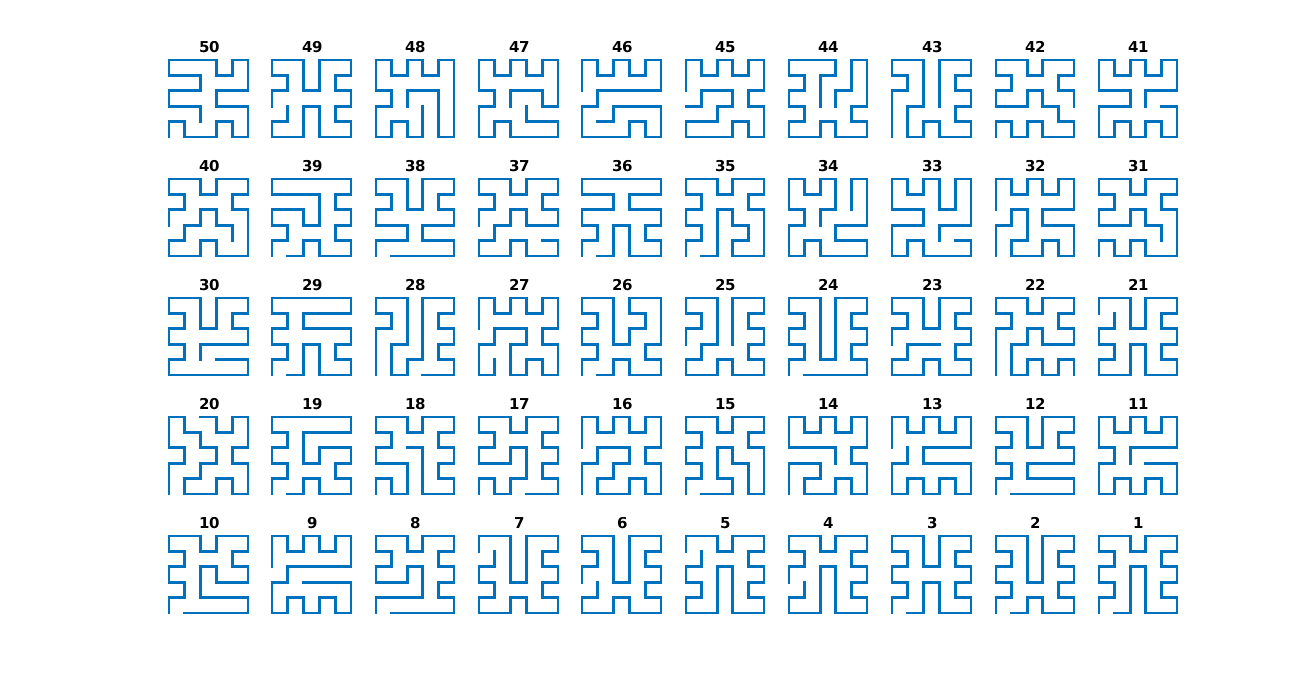}\\
\caption{50 most designable folds on the 6x6 lattice labelled with rank.}
\label{fig:6x6top}
\end{figure}

\begin{figure}[h!]
\centering
\includegraphics[width=0.8\linewidth]{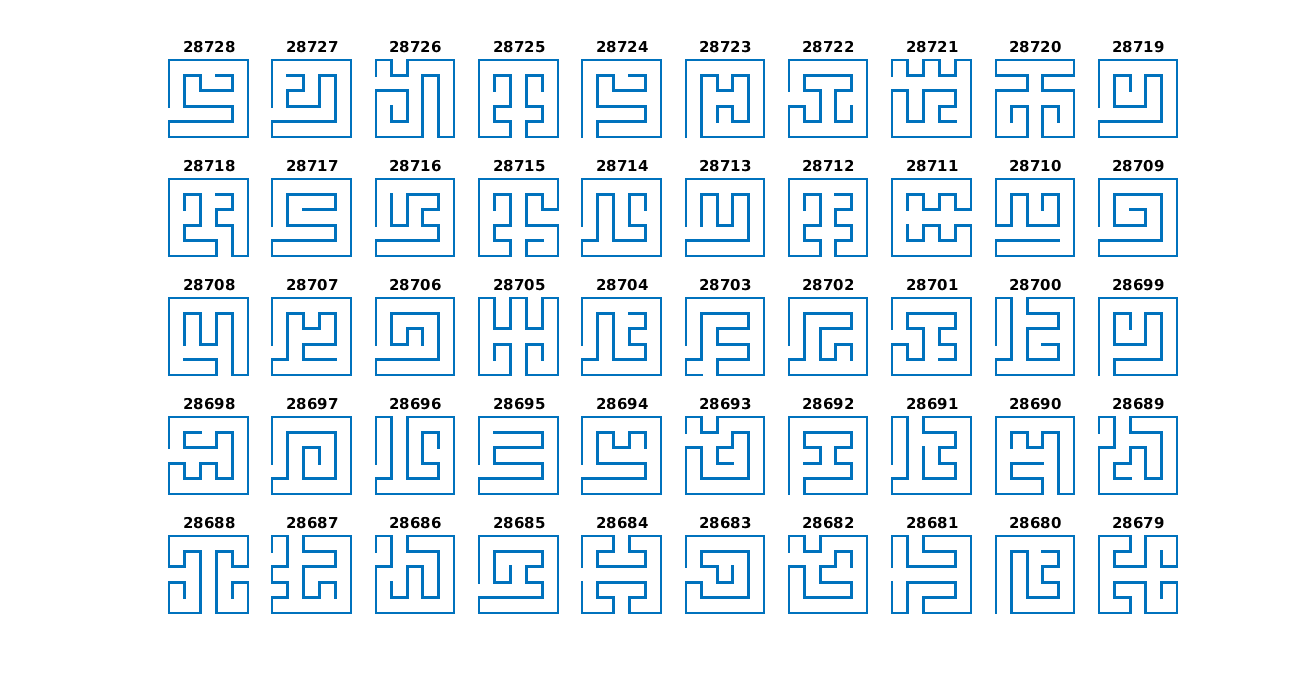}\\
\caption{50 least designable folds for the 6x6 lattice labelled with rank.}
\label{fig:6x6bot}
\end{figure}

The following folds on the 6x6 lattice were found to be the most (Figure \ref{fig:6x6top}) and least (Figure \ref{fig:6x6bot}) designable for the MJ1985 model on the 6x6 lattice. 

\clearpage
\section{Code}
\subsection{GPU Code}
\label{app:gpu}
GPU programs involve a compute kernel and the launching of that kernel from within the CPU code. The following code is the version for the HP1996 model on the 6x6 lattice.
Lines 27-29 define the number of blocks and threads per block. These parameters are used to launch the kernel and were set to the optimal values found in the results.
Lines 46-100 detail the kernel which computes and outputs the lowest energies. The most time consuming part for all random sampling versions is in lines 62-64 where the random sequence values are read in by each thread.
Line 215 is where the kernel is launched from the CPU.
\begin{lstlisting}
/*
CUDA ENERGY CALCULATION - RANDOM SAMPLING - SJOWEN - 20/03/15

In this code, for each sequence (counted in binary), the lowest energy fold is found, and if unique, the id of that fold is stored. The calculations are done reversing sign of the cross energy and the max energy / unique fold is found. 

HP1996 model for 6x6


*/


#include <time.h>
#include <stdio.h>
#include <stdlib.h>
#include <math.h>
#include <string>
#include <iostream>
#include <fstream>
#include <sstream>
#include <vector>
#include <cuda.h>
#include <curand.h>

#define N_FOLD		57337	 // number of 6x6 folds
#define length  	36	 // length sequence (from 3x3x3)
#define numseq  	512*8*4 // #threads = blocks * threadsperblock
#define dimx		512
#define dimy		8
#define thperblock	4	 // threadsperblock
#define threshold	1	 // min kT*10 between ground and 1st exc. 
#define NBP		25
#define runs		7

#define gpuErrchk(ans) { gpuAssert((ans), __FILE__, __LINE__); }
inline void gpuAssert(cudaError_t code, const char *file, int line, bool abort=true)
{
   if (code != cudaSuccess) 
   {
      fprintf(stderr,"GPUassert: %s %s %d\n", cudaGetErrorString(code), file, line);
      if (abort) exit(code);
   }
}



// cuda kernel 
// calculates foldid of lowest unique energy fold for a sequence that equals threadID
__global__ void energycalc(int V[N_FOLD][2*NBP], int SeqRnd[numseq][length], int *ID_fold, int *E_dif)
{

    int sequence[length];
    int blockId = blockIdx.x + blockIdx.y * gridDim.x;
    int threadId = blockId * blockDim.x + threadIdx.x; //global id of thread
    int min_E_sequence = 0;
    int min_E_sequence2 = 0;
    int E[4];
    int min_E_fid = -1;
    int energy_fold = 0;
    ID_fold[threadId]= -1;	// The ID of the min E fold.

    // fill sequence with random amino acids
    for (int jj=0;jj<length;jj++)  {
        sequence[jj]=SeqRnd[threadId][jj];
 	}   
  
    E[0]=-23;
    E[1]=-10;
    E[2]=-10;
    E[3]=0;
    
    // run through folds and find lowest energy
    for (int foldid = N_FOLD-1; foldid > -1; foldid--)
    {
        energy_fold=0;
	
       // k is less than half of neighbour partners
        for (int k=0;k<NBP;k++){ //iterate through neighbours for 
            int loc1 = V[foldid][2*k];
            int loc2 = V[foldid][2*k+1]; // location of neighbour
	    energy_fold = energy_fold + E[ sequence[loc1]+2*sequence[loc2] ];
	}
        
	// if energy of this fold is min 
    	
        if (energy_fold < min_E_sequence)  {
	     min_E_sequence2 = min_E_sequence;
	     min_E_fid = foldid;
	     min_E_sequence = energy_fold;
	}
	
	else if (energy_fold < min_E_sequence2) {
	     min_E_sequence2 = energy_fold;
        }

    }

	ID_fold[threadId] = min_E_fid;
	E_dif[threadId] = min_E_sequence - min_E_sequence2;
   
}


int main(void)
{
    std::vector< std::vector<int> > V;
    std::vector <int> vec;
    std::vector< std::vector<double> > VE;
    std::vector <double> vecE;
    int* energy = new int[numseq];
    typedef int T[2*NBP];
    typedef int S[length];
    int* minfold = new int[numseq];
    int* E_dif = new int[numseq];
    int* dev_foldID;
    T* dev_V = new T[N_FOLD];
    int* dev_Edif;
    T* V_array = new T[N_FOLD]; 
    S* SeqRnd = new S[numseq];
    S* dev_SeqRnd = new S[numseq];
    float pct;
    srand(time(NULL));

    // Fill V_array with -1s
    for (int n=0;n<N_FOLD;n++)
    {
	for (int m=0;m<2*NBP;m++)
	{	V_array[n][m]=-1;	
	}
    }	
	

    // allocate memory for V and min energy fold ids on GPU

    gpuErrchk(cudaMalloc( (void**)&dev_V, N_FOLD * 2*NBP * sizeof(int) ) );
    gpuErrchk(cudaMalloc( (void**)&dev_SeqRnd, numseq*length*sizeof(int) ));
    gpuErrchk(cudaMalloc( (void**)&dev_foldID, numseq*sizeof(int) ));
    gpuErrchk(cudaMalloc( (void**)&dev_Edif, numseq*sizeof(int) ));
    cudaGetLastError();

    // CPU code for initialising V vector of folds (list of neighbours)
    
    std::ifstream fin("data_neighbours_6x6.txt");
    std::string line;
    int i;
    while (std::getline( fin, line) ) //Read a line 
    {
       std::stringstream ss(line);

       while(ss >> i) {		//Extract integers from line of file
         vec.push_back(i);
       }
       
       V.push_back(vec); 
       vec.clear();
    }
    fin.close();
  
    
   // Copy from V to V_array
   for (int n=0;n<N_FOLD;n++)  { // Iterate over all folds
	for (int m=0;m<V[n].size();m++)	{ // Iterate over number of fold neighbour locations
		V_array[n][m]=V[n][m];	// convert read in vectors into array
	}	
   }
   
   V.clear();
   
   
    std::ifstream fin3("data_lookup_6x6.txt");
    std::string line3;
    std::vector <int> vec3;
    std::vector <std::vector <int> > V3;
    int i3;
    while (std::getline( fin3, line3) ) //Read a line 
    {
       std::stringstream ss(line3);

       while(ss >> i3) //Extract integers from line
         vec3.push_back(i3);

       V3.push_back(vec3); 
       vec3.clear();
    }
    fin.close();

    std::ofstream GSfolds;
    std::ofstream counter;
   
    // Copy V array of neighbour pairs to GPU, Random numbers for sequences and Energy matrix. 
    cudaMemcpy( dev_V,  V_array, N_FOLD*2*NBP*sizeof(int), cudaMemcpyHostToDevice ) ;
    
    clock_t t;
    
 
    
    for (int kk=0; kk<runs; kk++) {  
       dim3 blocks2(int(dimx/std::pow(2.0,kk)),dimy);
 
       counter << kk << std::endl;

    	GSfolds.open("GSfolds_HP1996_6x6.txt", std::ios::app);
	counter.open("count.txt", std::ios::app);

	t = clock();
    
	for (int n=0;n<numseq;n++)
        {
		for (int m=0;m<length;m++)
        	{       SeqRnd[n][m]=rand()%2;
        	}
    	}
    
	cudaMemcpy( dev_SeqRnd, SeqRnd, numseq*length*sizeof(int), cudaMemcpyHostToDevice );
    // Run GPU kernel
    energycalc<<<blocks2,thperblock*int(std::pow(2.0,kk))>>>(dev_V, dev_SeqRnd, dev_foldID, dev_Edif);
    
    // copy MAXE_fold ids back to host
    cudaMemcpy( minfold, dev_foldID, numseq*sizeof(int), cudaMemcpyDeviceToHost ) ;
    cudaMemcpy( E_dif, dev_Edif, numseq*sizeof(int), cudaMemcpyDeviceToHost ) ;
    t = clock() -t;
    std::cout << (float)t/(CLOCKS_PER_SEC*dimx*dimy*thperblock) << "s per fold." << std::endl;


    int ijk=0;
    // print out sequence id and fold id
    for (int zz=0; zz<numseq; zz++) {	
		GSfolds << V3[minfold[zz]][0] << " " ;
		GSfolds << E_dif[zz] << std::endl; 
		ijk++;
    }    
	
    double f = double(numseq);
    pct += ijk/f;
    
    std::cout << "Percentage: " << pct/runs << std::endl;
    counter.close();    
    GSfolds.close();
}

    //  Free memory on device

    cudaFree( dev_V );
    cudaFree( dev_SeqRnd );
    cudaFree( dev_Edif );
    cudaFree( dev_foldID );
    cudaGetLastError();
    
    return 0;
}

\end{lstlisting}

\end{appendices}

\end{document}